\documentclass[a4paper,11pt]{article}%
\usepackage{amsmath}
\usepackage{cite}
\usepackage{citesort}
\usepackage{graphicx}
\usepackage{amsbsy}%
\usepackage{amsfonts}%
\usepackage{amssymb}
\newcommand{\bm}[1]{\boldsymbol{#1}}

\begin{document}
\thispagestyle{empty}  \setcounter{page}{0}  \begin{flushright}%

October 2007 \\
\end{flushright}%

\vskip                                4.2 true cm

\begin{center}
{\huge Minimal Flavor Violation, Seesaw, and R-parity}\\[2.3cm]

\textsc{Emanuel Nikolidakis\footnote{nikolida@itp.unibe.ch} and Christopher
Smith\footnote{chsmith@itp.unibe.ch}}\\[12pt]\textsl{Institut f\"{u}r
Theoretische Physik, Universit\"{a}t Bern, CH-3012 Bern, Switzerland }\\[2.2cm]

\textbf{Abstract}
\end{center}

\begin{quote}
\noindent The Minimal Flavor Violation hypothesis (MFV) is extended to the
R-parity violating MSSM, supplemented with a simple seesaw mechanism. The
requirement of MFV is shown to suppress lepton and baryon-number violating
couplings sufficiently to pass all experimental bounds, in particular those
for proton decay, and is thus a viable alternative to R-parity. The
phenomenological consequences for FCNC, LFV and colliders are briefly
discussed. Typically, MFV predicts sizeable baryon-number violation in some
characteristic channels, like single stop resonant production.\newpage
\end{quote}

\section{Introduction}

The presence of the scalar partners of the quarks and leptons in the Minimal
Supersymmetric Standard Model (MSSM) directly allows for renormalizable
interactions violating baryon ($B$) or lepton ($L$) numbers. Though only
accidental in the Standard Model (SM), there is considerable experimental
support for the near conservation of at least one of these quantum numbers. In
particular, the current bound on the proton lifetime sets a very tough limit
on certain combinations of $\Delta L=1$ and $\Delta B=1$ couplings. To recover
a viable phenomenology, the MSSM incorporates R-parity\cite{FarrarF76}, which
explicitly forbids $\Delta L=1$ and $\Delta B=1$ interactions. At the same
time, this symmetry leads to a very distinctive phenomenology, and signatures
at colliders, since supersymmetric particles can be produced only in pairs,
and the lightest supersymmetric particle (LSP) is absolutely stable.

The MSSM is not thought to be the ultimate theory. In particular, to account
for the observed small neutrino masses by the seesaw mechanism\cite{SeeSaw},
its particle content has to be extended at some high-energy scale. In that
context, R-parity looses its main appeal since it no longer protects the
proton from decaying rapidly. Indeed, together with the $\Delta L=2$ Majorana
mass operator, R-parity conserving $\Delta L=1$, $\Delta B=1$ operators can
appear among the dimension-five effective interactions generated by
integrating out the high-energy degrees of freedom\cite{IbanezR91}. To resolve
this issue, many models have been proposed, which typically predict that
either $L$ or $B$ is exactly conserved at low-energy, but not necessarily
R-parity (for a review, see e.g. Ref.\cite{BarbierEtAl04}).

Besides the issue of proton stability, the presence of flavored scalar
particles seems at odds with the observed suppression of flavor-changing
processes, especially neutral currents and $CP$-violating phenomena. In the
SM, the CKM matrix, with its hierarchical structure, is able to account for
all experimental results with an impressive precision. Therefore, with squark
masses at or below the TeV to avoid destabilizing the electroweak scale, the
MSSM scalar sector must be precisely fine-tuned to preserve these delicate
patterns. Note that no fine-tuning is required in the supersymmetric sector,
where quarks and squarks are aligned, but it needs to be enforced by hand for
the soft-breaking terms, the remnants of the unknown supersymmetry-breaking mechanism.

A particularly elegant procedure to maintain this alignment, and thus to keep
squark-induced flavor-breakings in check, is to enforce it through a symmetry
principle, the so-called Minimal Flavor Violation (MFV) hypothesis of
Ref.\cite{DambrosioGIS02}. Though the origin of this symmetry is unclear, it
allows for precise and well-defined predictions. Further, if proven valid by
comparison with experiment, it may give us some glimpses of the mechanism at
the origin of the flavor structures, totally unexplained within the MSSM, and
will constrain the supersymmetry-breaking mechanism.

The MFV hypothesis is thus a systematic symmetry principle well-supported by
data. Our goal in the present work is to show that it can also explain proton
stability, without calling in any additional symmetry. To appreciate the
problem at hand, let us recall the orders of magnitude at play. If
baryon-number conservation is not enforced, and $\Delta L=2$ effects arise
through the seesaw mechanism, given the present limits on the proton lifetime
of more than $10^{30}$ years\cite{PDG}, our goal is to reconcile the following
magnitude estimates%
\begin{equation}
\lambda_{\Delta B}\sim\mathcal{O}(1),\;\lambda_{\Delta L=2}\sim m_{\nu
}/(100\,\text{GeV})\sim\mathcal{O}(10^{-12})\;\;\overset{{\Large ?}%
}{\rightarrow}\;|\lambda_{\Delta B}\lambda_{\Delta L=1}|\lesssim
\mathcal{O}(10^{-24})\;,
\end{equation}
for neutrino masses $m_{\nu}\lesssim1$ eV and sparticle masses below $1$ TeV.
Assuming that the $\Delta L=1$ and $\Delta L=2$ scales are similar, the
requirement of MFV alone will be seen to be able to bridge the remaining gap
of more than ten orders of magnitude, at least under certain circumstances.
Specifically, the suppression capabilities of MFV strongly depend on $m_{\nu}%
$, the ratio of the vacuum expectation values of the two MSSM Higgses
($\tan\beta\equiv v_{u}/v_{d}$), and on the flavor directions in which baryon
and lepton numbers are violated, a point which will be precisely defined later.

Finally, having replaced R-parity by MFV has many consequences. For example,
the LSP is able to decay, and single-sparticle events could be seen at
colliders. In that context, MFV predicts specific hierarchies among R-parity
violating couplings, and can thus tell us in which direction to look for
supersymmetric effects. Also, the study of the conditions under which MFV is
sufficient to stabilize the proton offers interesting indirect constraints on
other sectors, for example Flavor Changing Neutral Currents (FCNC) or Lepton
Flavor Violation (LFV) effects, which also depend on $\tan\beta$ or $m_{\nu}$.

The structure of the paper is as follows. In Section 2, we establish the
minimal spurion content, and apply the MFV principle to R-parity violating
couplings. The properties of the MFV expansions under redefinitions of the
Higgs and lepton fields are then analyzed, as well as the application to
higher-dimensional operators. The phenomenological consequences are explored
in Section 3. First, the MFV predictions for the order of magnitude of all
R-parity violating couplings are worked out. Then, the various bounds are
checked, in particular those from $\Delta B=1$ nucleon decay, and the
consequences for $\tan\beta$ and $m_{\nu}$ are discussed. Finally, the
consequences for supersymmetric signals at colliders are briefly analyzed, and
our results are summarized in the Conclusion.

\section{The MFV hypothesis for the R-parity violating MSSM}

In regard to their gauge interactions, the three generations of quarks and
leptons are decoupled, and physics is invariant under redefinitions of these
matter fields. This is the origin of the global $U(3)^{5}$ flavor
symmetry\cite{ChivukulaG87}.

In the MSSM, this symmetry is broken in many places. At the very least, it is
broken in the superpotential so as to reproduce the known SM flavor sector,
i.e. the quark and charged lepton masses and CKM mixings. A priori, the
soft-breaking squark and slepton masses and trilinear terms represent new
sources of flavor mixings\cite{Martin97}.

If R-parity is not enforced, there are additional $\Delta L=1$ and $\Delta
B=1$ couplings, both in the superpotential and among the soft-breaking terms
(see e.g. Ref.\cite{BarbierEtAl04} for a review). All these new couplings
break the $U(3)^{5}$ flavor symmetry, since at least the $U(1)$'s for the
lepton and baryon numbers do not survive.

This is not yet the full story. In the MSSM, there is no right-handed
neutrino, while the left-handed neutrinos are massless. Therefore, to fully
account for the leptonic flavor sector, i.e. neutrino masses and mixings,
additional sources of flavor-breaking must be introduced. In the present work,
neutrino masses are generated through a simple seesaw mechanism of type I,
from integrating out right-handed neutrinos at some high-energy
scale\cite{SeeSaw}.

There are thus many sources of breaking of the $U(3)^{5}$ symmetry. To reduce
them, the MFV principle is a very attractive hypothesis. Indeed, it starts
from the requirement of minimality: only the simplest breakings of the
$U(3)^{5}$ symmetry are allowed. By simplest is meant only the minimal sources
of breaking able to generate a realistic flavor sector, i.e. the known fermion
masses and mixings. Technically, these flavor breaking sources are
parametrized as spurions, i.e. non-dynamical fields in definite $U(3)^{5}$ representations.

To enforce MFV, the first step is to identify these elementary spurions, and
then to parametrize all the flavor-breaking sectors of the MSSM as invariants
under the flavor group. At that stage, the freedom to choose the directions in
which lepton and baryon-numbers are broken will play a special role. Indeed,
in practice, this freedom translates into the choice of which $\varepsilon
$-tensors among the numerical invariant tensors of the five $SU(3)\in
U(3)^{5}$ are to be used to construct invariants. The purpose of the present
section is to construct these MFV expansions, leaving numerical studies and
phenomenological discussions for Section 3.

\subsection{The seesaw mechanism and MFV spurions}

In the supersymmetric limit, and without R-parity violation, the MSSM
superpotential (denoting quark and lepton superfields as $Q=(u_{L},d_{L})^{T}%
$, $U=u_{R}^{\dagger}$, $D=d_{R}^{\dagger}$, $L=(\nu_{L},e_{L})^{T}$,
$E=e_{R}^{\dagger}$, generation indices by $I,J$, and the $SU(2)_{L}$ spinor
products by parentheses)%
\begin{equation}
W_{RPC}=U^{I}(\mathbf{Y}_{u})^{IJ}(Q^{J}H_{u})-D^{I}(\mathbf{Y}_{d}%
)^{IJ}(Q^{J}H_{d})-E^{I}(\mathbf{Y}_{\ell})^{IJ}(L^{J}H_{d})+\mu\left(
H_{u}H_{d}\right)  \;,\label{WRPC}%
\end{equation}
is the only source of breaking of the $U(3)^{5}$ flavor symmetry acting on the
(s)quarks and (s)leptons:%
\begin{equation}%
\begin{array}
[c]{cccccc}%
G_{f}= & \underbrace{SU\left(  3\right)  _{Q}\times SU\left(  3\right)
_{U}\times SU\left(  3\right)  _{D}} & \times & \underbrace{SU\left(
3\right)  _{L}\times SU\left(  3\right)  _{E}} & \times & G_{1}\;,\\
& G_{q} &  & G_{\ell} &  &
\end{array}
\end{equation}
with the $U\left(  1\right)  $'s acting on individual fields moved into the
group $G_{1}$. Upon rearranging them, it can be written as%
\begin{equation}
G_{1}=U\left(  1\right)  _{B}\times U\left(  1\right)  _{L}\times U\left(
1\right)  _{Y}\times U\left(  1\right)  _{PQ}\times U\left(  1\right)
_{E}\;.\label{U1}%
\end{equation}
The first three correspond to the conserved baryon number, lepton
number\footnote{We use the same notation for the lepton-number $U(1)_{L}$, and
for the $U(1)_{L}$ acting on the left-handed lepton doublet. In the following,
$U(1)_{L}$ always denotes the latter.} and weak hypercharge. The $U\left(
1\right)  _{PQ}$, acting on $D$ and $E$, is conserved if the Higgs $H_{d}$ is
also transforming non-trivially, and then is equivalent to the Peccei-Quinn
symmetry of the two Higgs doublet model\cite{PecceiQ77}. Finally, the
remaining $U(1)$ acts only on $E$ and is broken by the leptonic Yukawa coupling.

The superpotential $W_{RPC}$ is made formally invariant under $G_{f}$ by
promoting $\mathbf{Y}_{u,d,\ell}$ to spurion fields transforming
as\cite{DambrosioGIS02}%
\begin{gather}
U\overset{G_{f}}{\rightarrow}Ug_{U}^{\dagger},\;D\overset{G_{f}}{\rightarrow
}Dg_{D}^{\dagger},\;Q\overset{G_{f}}{\rightarrow}g_{Q}Q,\;E\overset{G_{f}%
}{\rightarrow}Eg_{E}^{\dagger},\;L\overset{G_{f}}{\rightarrow}g_{L}L,\;\\
\mathbf{Y}_{u}\overset{G_{f}}{\rightarrow}g_{U}\mathbf{Y}_{u}g_{Q}^{\dagger
},\;\mathbf{Y}_{d}\overset{G_{f}}{\rightarrow}g_{D}\mathbf{Y}_{d}%
g_{Q}^{\dagger},\;\mathbf{Y}_{\ell}\overset{G_{f}}{\rightarrow}g_{E}%
\mathbf{Y}_{\ell}g_{L}^{\dagger}, \label{Spurion1}%
\end{gather}
or $\mathbf{Y}_{u}\sim\left(  \bar{3},3,1\right)  _{G_{q}}$, $\mathbf{Y}%
_{d}\sim\left(  \bar{3},1,3\right)  _{G_{q}}$, $\mathbf{Y}_{\ell}\sim\left(
\bar{3},3\right)  _{G_{\ell}}$. Therefore, in order to account for the quark
masses, CKM mixings and charged lepton masses, one needs at least three
spurions with these transformation properties. The minimal case is when the
basic sources of flavor-breaking are only along the $\left(  \bar
{3},3,1\right)  _{G_{q}}$, $\left(  \bar{3},1,3\right)  _{G_{q}}$ and $\left(
\bar{3},3\right)  _{G_{\ell}}$ directions.

There remain the light neutrino masses. For them, we supplement the MSSM with
a seesaw mechanism\cite{SeeSaw}, following Ref.\cite{CiriglianoGIW05}. We
start by adding heavy right-handed (s)neutrinos:%
\begin{equation}
W_{\nu_{R}}=W_{RPC}+\frac{1}{2}N^{I}\mathbf{M}^{IJ}N^{J}+N^{I}(\mathbf{Y}%
_{\nu})^{IJ}(L^{J}H_{u})\;,
\end{equation}
corresponding to an enlarged flavor-symmetry $G_{f}\times U\left(  3\right)
_{N}$ at the high-energy scale, with%
\begin{equation}
N\overset{G_{f}\times U\left(  3\right)  _{N}}{\rightarrow}Ng_{N}^{\dagger
},\;\;\mathbf{Y}_{\nu}\overset{G_{f}\times U\left(  3\right)  _{N}%
}{\rightarrow}g_{N}\mathbf{Y}_{\nu}g_{L}^{\dagger}\;,\;\;\mathbf{M}%
\overset{G_{f}\times U\left(  3\right)  _{N}}{\rightarrow}g_{N}\mathbf{M}%
g_{N}^{T}\;.
\end{equation}
When the Majorana mass $\mathbf{M}$ is very large, right-handed neutrinos can
be integrated out, leading to the well-known non-renormalizable dimension-five
term in the superpotential\cite{Weinberg79}%
\begin{equation}
W_{\dim-5}=\frac{1}{2}(\mathbf{Y}_{\nu})^{IK}(L^{K}H_{u})\left(
\mathbf{M}^{-1}\right)  ^{IJ}(\mathbf{Y}_{\nu})^{JL}(L^{L}H_{u})\overset
{\text{SSB}}{\rightarrow}\frac{1}{2}v_{u}^{2}\nu_{L}^{I}(\mathbf{Y}_{\nu
})^{IK}\left(  \mathbf{M}^{-1}\right)  ^{IJ}(\mathbf{Y}_{\nu})^{JL}\nu_{L}%
^{L}\;. \label{Dim5nu}%
\end{equation}
This term gives mass to the left-handed neutrinos after electroweak symmetry
breaking, and we define the dimensionless neutrino mass spurion as%
\begin{equation}
\mathbf{\Upsilon}_{\nu}\equiv v_{u}\mathbf{Y}_{\nu}^{T}\mathbf{M}%
^{-1}\mathbf{Y}_{\nu}\sim\mathcal{O}(m_{\nu}/v_{u})\;,
\end{equation}
transforming as $\mathbf{\Upsilon}_{\nu}\sim\left(  \bar{6},1\right)
_{G_{\ell}}$. It is symmetric, $\mathbf{\Upsilon}_{\nu}=\mathbf{\Upsilon}%
_{\nu}^{T}$, since $\mathbf{M}^{-1}$ can be assumed diagonal without loss of
generality. For simplicity, we further assume
\begin{equation}
\mathbf{M}=M_{R}\mathbf{1\;.}%
\end{equation}
This is the well-known seesaw mechanism: the heavy mass scale $M_{R}$ bears
the responsibility for the smallness of the neutrino masses, not the Yukawa
$\mathbf{Y}_{\nu}$. For example, with $m_{\nu}\sim1$ eV, $\mathbf{Y}_{\nu}%
\sim\mathcal{O}(1)$ when $M_{R}\sim10^{13}$ GeV.

For consistency, we must include also the other spurion transforming as a
singlet under $U\left(  3\right)  _{N}$, which is $\mathbf{Y}_{\nu}^{\dagger
}\mathbf{Y}_{\nu}\sim\left(  8,1\right)  _{G_{\ell}}$. Compared to
$\mathbf{\Upsilon}_{\nu}$, it is not suppressed by the heavy mass scale. It
has to be included into our list of spurions because it is transforming
differently than $\mathbf{\Upsilon}_{\nu}$. Further, it plays an important
role in the generation of LFV effects if the supersymmetry-breaking scale is
much higher than the $M_{R}$ scale\cite{BorzumatiM86}. Therefore, the two
neutrino spurions which have to be included are%
\begin{equation}
\mathbf{\Upsilon}_{\nu}\overset{G_{f}}{\rightarrow}g_{L}^{\ast}%
\mathbf{\Upsilon}_{\nu}g_{L}^{\dagger},\;\;\mathbf{Y}_{\nu}^{\dagger
}\mathbf{Y}_{\nu}\overset{G_{f}}{\rightarrow}g_{L}\mathbf{Y}_{\nu}^{\dagger
}\mathbf{Y}_{\nu}g_{L}^{\dagger}\;. \label{EFC}%
\end{equation}

What is important about the spurions in Eqs.(\ref{Spurion1}) and (\ref{EFC})
is their specific transformation properties under the flavor group, not their
expressions in terms of the Yukawas. For instance, consider the $U(QH_{u})$
term of $W_{RPC}$. As soon as the flavor symmetry is broken along the $\left(
\bar{3},3,1\right)  _{G_{q}}$ and $\left(  \bar{3},1,3\right)  _{G_{q}}$
directions, which we parametrize by the two spurions $\mathbf{Y}_{u}$ and
$\mathbf{Y}_{d}$, MFV implies that we can write%
\begin{equation}
W_{RPC}=U^{I}(a_{1}\mathbf{Y}_{u}+a_{2}\mathbf{Y}_{u}\mathbf{Y}_{u}^{\dagger
}\mathbf{Y}_{u}+a_{3}\mathbf{Y}_{u}\mathbf{Y}_{d}^{\dagger}\mathbf{Y}%
_{d}...)^{IJ}(Q^{J}H_{u})+\,...\;,
\end{equation}
with MFV coefficients $a_{i}\sim\mathcal{O}(1)$. Since it is always possible
to redefine the spurions as $a_{1}\mathbf{Y}_{u}+a_{2}\mathbf{Y}_{u}%
\mathbf{Y}_{u}^{\dagger}\mathbf{Y}_{u}+a_{3}\mathbf{Y}_{u}\mathbf{Y}%
_{d}^{\dagger}\mathbf{Y}_{d}...\rightarrow\mathbf{Y}_{u}$, which is neutral
from the point of view of $G_{f}$, the Yukawa couplings can be assumed to take
their usual forms, and there is no need to distinguish between spurions fields
and Yukawas couplings. In addition, using the freedom to perform flavor
rotations, they can be brought to their background values%
\begin{equation}
\mathbf{Y}_{u}=\mathbf{m}_{u}V/v_{u},\;\;\mathbf{Y}_{d}=\mathbf{m}_{d}%
/v_{d},\;\;\mathbf{Y}_{\ell}=\mathbf{m}_{\ell}/v_{d},\;\;\mathbf{\Upsilon
}_{\nu}=U^{\ast}\mathbf{m}_{\nu}U^{\dagger}/v_{u},\;\;\mathbf{Y}_{\nu
}^{\dagger}\mathbf{Y}_{\nu}\overset{CP}{=}M_{R}\mathbf{\Upsilon}_{\nu
}\mathbf{/}v_{u}\;, \label{Background}%
\end{equation}
where $V$ is the CKM matrix, $U$ the PMNS matrix, $\mathbf{m}_{u,d,\ell,\nu}$
are the diagonal fermion mass matrices and $v_{u,d}$ the VEV's of the
$H_{u,d}^{0}$ Higgs, with $v_{u}^{2}+v_{d}^{2}\approx(174\;$GeV)$^{2}$ and
$\tan\beta\equiv v_{u}/v_{d}$. The spurion $\mathbf{Y}_{\nu}^{\dagger
}\mathbf{Y}_{\nu}$ can be fixed only in the $CP$-limit, by neglecting unknown
phases (see Section 3.1). It is important to stress that we only take the
$CP$-limit in numerical estimates. For the purpose of enforcing the MFV
hypothesis, it is essential to keep track of their different transformation
properties under $SU(3)_{L}$, namely $\mathbf{Y}_{\nu}^{\dagger}%
\mathbf{Y}_{\nu}\sim\left(  8,1\right)  _{G_{\ell}}$ and $\mathbf{\Upsilon
}_{\nu}\sim\left(  \bar{6},1\right)  _{G_{\ell}}$.

\subsection{The general MFV expansion}

In the previous section, we established the minimal set of spurions needed to
generate a phenomenologically viable quark and lepton flavor-breaking sector.
Out of them, we now parametrize all the other flavor-breaking sectors of the
MSSM, including both R-parity conserving ($RPC$) and R-parity violating
($RPV$) couplings (see e.g. Ref.\cite{BarbierEtAl04} for a review).
Specifically, the $RPV$ superpotential terms are%
\begin{equation}
W_{RPV}=\frac{1}{2}%
%TCIMACRO{\TeXButton{lambda}{\bm{\lambda}}}%
%BeginExpansion
\bm{\lambda}%
%EndExpansion
^{IJK}(L^{I}L^{J})E^{K}+%
%TCIMACRO{\TeXButton{lambda}{\bm{\lambda}}}%
%BeginExpansion
\bm{\lambda}%
%EndExpansion
^{\prime IJK}(L^{I}Q^{J})D^{K}+%
%TCIMACRO{\TeXButton{mu}{\bm{\mu}}}%
%BeginExpansion
\bm{\mu}%
%EndExpansion
^{\prime I}(H_{u}L^{I})+\frac{1}{2}%
%TCIMACRO{\TeXButton{lambda}{\bm{\lambda}}}%
%BeginExpansion
\bm{\lambda}%
%EndExpansion
^{\prime\prime IJK}U^{I}D^{J}D^{K}\;, \label{WRPV}%
\end{equation}
while the $RPC$ and $RPV$ soft-breaking terms involving the scalar fields are%
\begin{subequations}
\begin{align}
\mathcal{L}_{\text{soft}}^{RPC-bilinear}  &  =-m_{H_{u}}^{2}H_{u}^{\dagger
}H_{u}-m_{H_{d}}^{2}H_{d}^{\dagger}H_{d}-(b(H_{u}H_{d})+h.c.)\frac{{}}{{}%
}\nonumber\\
&  \;\;\;\;-\tilde{Q}^{\dagger}\mathbf{m}_{Q}^{2}\cdot\tilde{Q}%
-\tilde{U}\mathbf{m}_{U}^{2}\tilde{U}^{\dagger}-\tilde{D}\mathbf{m}_{D}%
^{2}\tilde{D}^{\dagger}-\tilde{L}^{\dagger}\mathbf{m}_{L}^{2}\cdot\tilde
{L}-\tilde{E}\mathbf{m}_{E}^{2}\tilde{E}^{\dagger}\;,\frac{{}}{{}%
}\label{RPCbi}\\
\mathcal{L}_{\text{soft}}^{RPC-trilinear}  &  =-\tilde{U}^{I}(\mathbf{A}%
_{u})^{IJ}(\tilde{Q}^{J}H_{u})+\tilde{D}^{I}(\mathbf{A}_{d})^{IJ}(\tilde
{Q}^{J}H_{d})+\tilde{E}^{I}(\mathbf{A}_{\ell})^{IJ}(\tilde{L}^{J}%
H_{d})+h.c.\;,\frac{{}}{{}}\label{RPCtri}\\
\mathcal{L}_{\text{soft}}^{RPV-bilinear}  &  =-\mathbf{b}^{\prime I}%
(H_{u}\tilde{L}^{I})-(\mathbf{m}_{Ld}^{2})^{I}H_{d}^{\dagger}\cdot\tilde
{L}^{I}+h.c.\;,\frac{{}}{{}}\label{RPVbi}\\
\mathcal{L}_{\text{soft}}^{RPV-trilinear}  &  =\frac{1}{2}\mathbf{A}%
^{IJK}(\tilde{L}^{I}\tilde{L}^{J})\tilde{E}^{K}+\mathbf{A}^{\prime IJK}%
(\tilde{L}^{I}\tilde{Q}^{J})\tilde{D}^{K}+\frac{1}{2}\mathbf{A}^{\prime\prime
IJK}\tilde{U}^{I}\tilde{D}^{J}\tilde{D}^{K}+h.c.\;, \label{RPVtri}%
\end{align}
with the trilinear terms $%
%TCIMACRO{\TeXButton{lambda}{\bm{\lambda}}}%
%BeginExpansion
\bm{\lambda}%
%EndExpansion
^{IJK}$ and $\mathbf{A}^{IJK}$ ($%
%TCIMACRO{\TeXButton{lambda}{\bm{\lambda}}}%
%BeginExpansion
\bm{\lambda}%
%EndExpansion
^{\prime\prime IJK}$ and $\mathbf{A}^{\prime\prime IJK}$) antisymmetric under
$I\leftrightarrow J$ ($J\leftrightarrow K$).

The MFV hypothesis is enforced by making all these couplings invariant under
$G_{f}$, up to the $U(1)$'s which are a priori broken, using only the
available spurions. As said before, there is no loss of generality in
identifying these spurions such that Eq.(\ref{Background}) holds, since this
corresponds to a mere $\mathcal{O}(1)$ redefinition for the MFV coefficients.

\subsubsection{Algebraic reductions and application to $RPC$ soft-breaking terms}

To get overall singlets under $G_{q}\times G_{\ell}$, the two invariant
tensors of the five $SU(3)$ can be used, namely $\delta^{IJ}$ and
$\varepsilon^{IJK}$. Given the large number of possible terms, we proceed in steps.

Let us first consider the two generic terms transforming as left-handed
octets, $\mathbf{R}_{\ell}\sim\left(  8,1\right)  _{G_{\ell}}$ and
$\mathbf{R}_{q}\sim\left(  8,1,1\right)  _{G_{q}}$. Making use of the
Cayley-Hamilton relation%
\end{subequations}
\begin{equation}
\mathbf{A}^{3}-Tr(\mathbf{A)A}^{2}+\frac{1}{2}\mathbf{A}(Tr(\mathbf{A)}%
^{2}-Tr(\mathbf{A}^{2}))-\frac{1}{3}Tr(\mathbf{A}^{3})+\frac{1}{2}%
Tr(\mathbf{A})Tr(\mathbf{A}^{2})-\frac{1}{6}Tr(\mathbf{A})^{3}=0\;, \label{CH}%
\end{equation}
together with the third-generation dominance (valid to about $5\%$)%
\begin{equation}
(\mathbf{Y}_{u,d,\ell}^{\dagger}\mathbf{Y}_{u,d,\ell})^{2}\approx y_{t,b,\tau
}^{2}\mathbf{Y}_{u,d,\ell}^{\dagger}\mathbf{Y}_{u,d,\ell}\;, \label{Num1}%
\end{equation}
with $y_{t}=m_{t}/v_{u}$ and $y_{b,\tau}=m_{b,\tau}/v_{d}$, reduces the number
of relevant octet terms to%
\begin{align}
\mathbf{R}_{q}  &  =\mathbf{1},\;\mathbf{Y}_{u}^{\dagger}\mathbf{Y}%
_{u},\;\mathbf{Y}_{d}^{\dagger}\mathbf{Y}_{d},\;\mathbf{Y}_{d}^{\dagger
}\mathbf{Y}_{d}\mathbf{Y}_{u}^{\dagger}\mathbf{Y}_{u},\;\mathbf{Y}%
_{u}^{\dagger}\mathbf{Y}_{u}\mathbf{Y}_{d}^{\dagger}\mathbf{Y}_{d}%
\;,\label{Xq}\\
\mathbf{R}_{\ell}  &  =\mathbf{1},\;\mathbf{Y}_{\ell}^{\dagger}\mathbf{Y}%
_{\ell},\;\mathbf{Y}_{\nu}^{\dagger}\mathbf{Y}_{\nu},\;\mathbf{Y}_{\nu
}^{\dagger}\mathbf{Y}_{\nu}\mathbf{Y}_{\ell}^{\dagger}\mathbf{Y}_{\ell
},\;\mathbf{Y}_{\ell}^{\dagger}\mathbf{Y}_{\ell}\mathbf{Y}_{\nu}^{\dagger
}\mathbf{Y}_{\nu},\;(\mathbf{Y}_{\nu}^{\dagger}\mathbf{Y}_{\nu})^{2}%
,\nonumber\\
&  \;\;\;\;\mathbf{Y}_{\ell}^{\dagger}\mathbf{Y}_{\ell}(\mathbf{Y}_{\nu
}^{\dagger}\mathbf{Y}_{\nu})^{2},\;(\mathbf{Y}_{\nu}^{\dagger}\mathbf{Y}_{\nu
})^{2}\mathbf{Y}_{\ell}^{\dagger}\mathbf{Y}_{\ell},\;(\mathbf{Y}_{\nu
}^{\dagger}\mathbf{Y}_{\nu})^{2}\mathbf{Y}_{\ell}^{\dagger}\mathbf{Y}_{\ell
}\mathbf{Y}_{\nu}^{\dagger}\mathbf{Y}_{\nu}\;. \label{Xl}%
\end{align}
We also used identities involving two or more different matrices to reach this
minimal basis. They can be found from Eq.(\ref{CH}) by expressing
$\mathbf{A}=a_{1}\mathbf{A}_{1}+a_{2}\mathbf{A}_{2}+...$ and extracting a
given power of $a_{1}$, $a_{2},...$. Importantly, since the $y_{t,b,\tau}^{2}$
are at most of $\mathcal{O}(1)$, these identities as well as Eq.(\ref{Num1})
do not generate large numerical coefficients.

The spurion $\mathbf{\Upsilon}_{\nu}$ was not used because it is very
suppressed compared to the others. Also, there is no need to consider
contractions with $\varepsilon$-tensors. Indeed, all such terms necessarily
involve an even number of $\varepsilon$-tensors, which can be simplified to
products or determinants of $\mathbf{R}_{i}$ monomials using%
\begin{equation}
\varepsilon^{IJK}\varepsilon^{LMN}=\det\left(
\begin{array}
[c]{ccc}%
\delta^{IL} & \delta^{IM} & \delta^{IN}\\
\delta^{JL} & \delta^{JM} & \delta^{JN}\\
\delta^{KL} & \delta^{KM} & \delta^{KN}%
\end{array}
\right)  ,\;\;\varepsilon^{LMN}\mathbf{A}^{LI}\mathbf{A}^{MJ}\mathbf{A}%
^{NK}=\det(\mathbf{A})\varepsilon^{IJK}\;. \label{epsi}%
\end{equation}

The next step is to construct the skeleton decompositions of each coupling,
and to dress them with all possible insertions of $\mathbf{R}_{q}$ and
$\mathbf{R}_{\ell}$ monomials. For example, looking at $\mathbf{m}_{U}^{2}$,
it transforms as $\mathbf{m}_{U}^{2}\rightarrow g_{U}\mathbf{m}_{U}^{2}%
g_{U}^{\dagger}$, hence its skeleton is $\mathbf{m}_{U}^{2}=m_{0}^{2}%
(a_{1}\mathbf{1}+a_{2}\mathbf{Y}_{u}\mathbf{Y}_{u}^{\dagger})$ with MFV
coefficients $a_{i}\sim\mathcal{O}(1)$ and $m_{0}^{2}$ setting the
supersymmetry breaking scale, as in $mSUGRA$\cite{Martin97}. All the relevant
MFV terms are then found by inserting $\mathbf{R}_{q}$ monomials between
$\mathbf{Y}_{u}$ and $\mathbf{Y}_{u}^{\dagger}$. With the further requirement
of hermicity for $\mathbf{m}_{U}^{2}$, we find%
\begin{equation}
\mathbf{m}_{U}^{2}=m_{0}^{2}(a_{1}\mathbf{1}+\mathbf{Y}_{u}(a_{2}%
\mathbf{1}+a_{3}\mathbf{Y}_{u}^{\dagger}\mathbf{Y}_{u}+a_{4}\mathbf{Y}%
_{d}^{\dagger}\mathbf{Y}_{d}+a_{5}(\mathbf{Y}_{d}^{\dagger}\mathbf{Y}%
_{d}\mathbf{Y}_{u}^{\dagger}\mathbf{Y}_{u}+\mathbf{Y}_{u}^{\dagger}%
\mathbf{Y}_{u}\mathbf{Y}_{d}^{\dagger}\mathbf{Y}_{d}))\mathbf{Y}_{u}^{\dagger
})\;,
\end{equation}
with MFV coefficients $a_{i}\sim\mathcal{O}(1)$. For compactness, we denote
this expansion as $\mathbf{m}_{U}^{2}=m_{0}^{2}(\mathbf{1}+\mathbf{Y}%
_{u}[\mathbf{R}_{q}]_{hc}\mathbf{Y}_{u}^{\dagger})$, where [...]$_{hc}$ stands
for the hermitian combination.

Proceeding similarly for the other $RPC$ soft-breaking terms, we find, written
in the compact form (arbitrary $\mathcal{O}(1)$ MFV coefficients are
understood everywhere)\cite{InPrep}:
\begin{subequations}
\label{RPCsoft}%
\begin{gather}
\mathbf{m}_{Q}^{2}=m_{0}^{2}[\mathbf{R}_{q}]_{hc},\;\;\mathbf{m}_{U}^{2}%
=m_{0}^{2}(\mathbf{1}+\mathbf{Y}_{u}[\mathbf{R}_{q}]_{hc}\mathbf{Y}%
_{u}^{\dagger}),\;\;\mathbf{m}_{D}^{2}=m_{0}^{2}(\mathbf{1}+\mathbf{Y}%
_{d}[\mathbf{R}_{q}]_{hc}\mathbf{Y}_{d}^{\dagger}),\frac{{}}{{}}\\
(\mathbf{A}_{u})^{IJ}=A_{0}((\mathbf{Y}_{u}\mathbf{R}_{q})^{IJ}+\varepsilon
^{LMN}\varepsilon^{ABC}(\mathbf{R}_{u})^{IA}(\mathbf{R}_{q}\mathbf{Y}%
_{u}^{\dagger})^{LB}(\mathbf{R}_{q}\mathbf{Y}_{u}^{\dagger})^{MC}%
(\mathbf{R}_{q})^{NJ}),\frac{{}}{{}}\\
(\mathbf{A}_{d})^{IJ}=A_{0}((\mathbf{Y}_{d}\mathbf{R}_{q})^{IJ}+\varepsilon
^{LMN}\varepsilon^{ABC}(\mathbf{R}_{d})^{IA}(\mathbf{R}_{q}\mathbf{Y}%
_{d}^{\dagger})^{LB}(\mathbf{R}_{q}\mathbf{Y}_{d}^{\dagger})^{MC}%
(\mathbf{R}_{q})^{NJ}),\frac{{}}{{}}\\
\mathbf{m}_{L}^{2}=m_{0}^{2}[\mathbf{R}_{\ell}]_{hc},\;\;\mathbf{m}_{E}%
^{2}=m_{0}^{2}(\mathbf{1}+\mathbf{Y}_{\ell}[\mathbf{R}_{\ell}]_{hc}%
\mathbf{Y}_{\ell}^{\dagger}),\frac{{}}{{}}\\
(\mathbf{A}_{\ell})^{IJ}=A_{0}((\mathbf{Y}_{\ell}\mathbf{R}_{\ell}%
)^{IJ}+\varepsilon^{LMN}\varepsilon^{ABC}(\mathbf{R}_{e})^{IA}(\mathbf{R}%
_{\ell}\mathbf{Y}_{\ell}^{\dagger})^{LB}(\mathbf{R}_{\ell}\mathbf{Y}_{\ell
}^{\dagger})^{MC}(\mathbf{R}_{\ell})^{NJ}),\frac{{}}{{}}%
\end{gather}
where the $\varepsilon$-terms have been reduced using Eq.(\ref{epsi}) as long
as only manifestly invariant terms under $G_{f}$ are generated, and the
$SU(3)_{U}$, $SU(3)_{D}$ and $SU(3)_{E}$ octets are defined as%
\end{subequations}
\begin{equation}
\mathbf{R}_{u}=\mathbf{1}+\mathbf{Y}_{u}\mathbf{R}_{q}\mathbf{Y}_{u}^{\dagger
},\;\mathbf{R}_{d}=\mathbf{1}+\mathbf{Y}_{d}\mathbf{R}_{q}\mathbf{Y}%
_{d}^{\dagger},\;\mathbf{R}_{e}=\mathbf{1}+\mathbf{Y}_{\ell}\mathbf{R}_{\ell
}\mathbf{Y}_{\ell}^{\dagger}\;, \label{Rright}%
\end{equation}
where, again, arbitrary $\mathcal{O}(1)$ coefficients are understood everywhere.

Apart from the additional $\varepsilon$-structures for the trilinear terms
$\mathbf{A}_{u,d}$, these expansions agree with those of
Refs.\cite{HallR90,DambrosioGIS02}, and their phenomenological consequences
for FCNC were analyzed in Refs.\cite{Pheno}. In the leptonic sector, to our
knowledge, they have not yet been written in this form. Note that
$\mathbf{\Upsilon}_{\nu}$ does not enter these expansions, and all LFV effects
arise from the non-diagonal $\mathbf{Y}_{\nu}^{\dagger}\mathbf{Y}_{\nu}$
spurion, as in Ref.\cite{BorzumatiM86}.

Concerning the $\varepsilon$-terms, their presence is unavoidable if one
sticks to the MFV principle. Though it is clear that they cannot emerge from
the RGE evolution of universal soft-breakings terms in the $RPC$ MSSM, they
are in general allowed once $RPV$ couplings are introduced. Since their
relevance for phenomenology has not yet been investigated, it is worth to
briefly describe their structure, leaving a detailed study for future work.
The $\varepsilon$-tensors being antisymmetric, their contributions have an
inverted hierarchy compared to the Yukawa terms, and are always small,
proportional to light-fermion masses. Anticipating on the results of the
numerical analyses of Section 3, only $(\mathbf{A}_{d})^{11}$ is significantly
affected by the $\varepsilon$-terms, and this only at large $\tan\beta$ (see
Appendix B).%

%TCIMACRO{\TeXButton{B}{\begin{table}[t] \centering}}%
%BeginExpansion
\begin{table}[t] \centering
%EndExpansion
$%
\begin{tabular}
[c]{lll}\hline
Struct. & $\text{MFV terms}$ & Broken $U(1)$\\\hline
\multicolumn{1}{c}{$%
%TCIMACRO{\TeXButton{mu}{\bm{\mu}}}%
%BeginExpansion
\bm{\mu}%
%EndExpansion
_{1}^{\prime I}$\rule{0in}{0.17in}} & $\mu\mathbf{\bar{\Upsilon}}_{\nu}%
^{I}\smallskip$ & \multicolumn{1}{c}{$\det(g_{L})$}\\\hline
\multicolumn{1}{c}{$%
%TCIMACRO{\TeXButton{lambda}{\bm{\lambda}}}%
%BeginExpansion
\bm{\lambda}%
%EndExpansion
_{1}^{IJK}$\rule{0in}{0.17in}} & $\mathbf{\bar{\Upsilon}}_{\nu}^{I}%
(\mathbf{Y}_{\ell}\mathbf{R}_{\ell})^{KJ}\smallskip$ &
\multicolumn{1}{c}{$\det(g_{L})$}\\
\multicolumn{1}{c}{$%
%TCIMACRO{\TeXButton{lambda}{\bm{\lambda}}}%
%BeginExpansion
\bm{\lambda}%
%EndExpansion
_{2}^{IJK}$\rule{0in}{0.17in}} & $\varepsilon^{LMN}(\mathbf{R}_{\ell}%
)^{LI}(\mathbf{Y}_{\ell}\mathbf{R}_{\ell}\mathbf{\Upsilon}_{\nu}^{\dagger
}\mathbf{R}_{\ell}^{T})^{KM}(\mathbf{R}_{\ell})^{NJ}\smallskip$ &
\multicolumn{1}{c}{$\det(g_{L})$}\\
\multicolumn{1}{c}{$%
%TCIMACRO{\TeXButton{lambda}{\bm{\lambda}}}%
%BeginExpansion
\bm{\lambda}%
%EndExpansion
_{3}^{IJK}$\rule{0in}{0.17in}} & $\mathbf{\bar{\Upsilon}}_{\nu}^{I}%
\varepsilon^{LMN}\varepsilon^{ABC}(\mathbf{R}_{e})^{KA}(\mathbf{R}_{\ell
}\mathbf{Y}_{\ell}^{\dagger})^{LB}(\mathbf{R}_{\ell}\mathbf{Y}_{\ell}%
^{\dagger})^{MC}(\mathbf{R}_{\ell})^{NJ}\smallskip$ & \multicolumn{1}{c}{$\det
(g_{L}^{2}g_{E}^{\dagger})$}\\
\multicolumn{1}{c}{$%
%TCIMACRO{\TeXButton{lambda}{\bm{\lambda}}}%
%BeginExpansion
\bm{\lambda}%
%EndExpansion
_{4}^{IJK}$\rule{0in}{0.17in}} & $\varepsilon^{LMN}\mathbf{\epsilon}%
^{ABCDEF}(\mathbf{R}_{\ell}\mathbf{Y}_{\ell}^{\dagger})^{AL}(\mathbf{R}_{\ell
}\mathbf{Y}_{\ell}^{\dagger})^{BM}(\mathbf{R}_{e})^{KN}(\mathbf{R}_{\ell
}\mathbf{\Upsilon}_{\nu}^{\dagger}\mathbf{R}_{\ell}^{T})^{CD}(\mathbf{R}%
_{\ell})^{EI}(\mathbf{R}_{\ell})^{FJ}\smallskip$ & \multicolumn{1}{c}{$\det
(g_{L}^{2}g_{E}^{\dagger})$}\\\hline
\multicolumn{1}{c}{$%
%TCIMACRO{\TeXButton{lambda}{\bm{\lambda}}}%
%BeginExpansion
\bm{\lambda}%
%EndExpansion
_{1}^{\prime IJK}$\rule{0in}{0.17in}} & $\mathbf{\bar{\Upsilon}}_{\nu}%
^{I}(\mathbf{Y}_{d}\mathbf{R}_{q})^{KJ}\smallskip$ & \multicolumn{1}{c}{$\det
(g_{L})$}\\
\multicolumn{1}{c}{$%
%TCIMACRO{\TeXButton{lambda}{\bm{\lambda}}}%
%BeginExpansion
\bm{\lambda}%
%EndExpansion
_{2}^{\prime IJK}$\rule{0in}{0.17in}} & $\mathbf{\bar{\Upsilon}}_{\nu}%
^{I}\varepsilon^{LMN}\varepsilon^{ABC}(\mathbf{R}_{d})^{KA}(\mathbf{R}%
_{q}\mathbf{Y}_{d}^{\dagger})^{LB}(\mathbf{R}_{q}\mathbf{Y}_{d}^{\dagger
})^{MC}(\mathbf{R}_{q})^{NJ}\smallskip$ & \multicolumn{1}{c}{$\det(g_{L}%
g_{D}^{\dagger}g_{Q})$}\\\hline
\multicolumn{1}{c}{$%
%TCIMACRO{\TeXButton{lambda}{\bm{\lambda}}}%
%BeginExpansion
\bm{\lambda}%
%EndExpansion
_{1}^{\prime\prime IJK}$\rule{0in}{0.17in}} & $\varepsilon^{LMN}%
(\mathbf{Y}_{u}\mathbf{R}_{q}\mathbf{Y}_{d}^{\dagger})^{IL}(\mathbf{R}%
_{d})^{JM}(\mathbf{R}_{d})^{KN}\smallskip$ & \multicolumn{1}{c}{$\det
(g_{D}^{\dagger})$}\\
\multicolumn{1}{c}{$%
%TCIMACRO{\TeXButton{lambda}{\bm{\lambda}}}%
%BeginExpansion
\bm{\lambda}%
%EndExpansion
_{2}^{\prime\prime IJK}$\rule{0in}{0.17in}} & $\varepsilon^{LMN}%
(\mathbf{R}_{u})^{IL}(\mathbf{Y}_{d}\mathbf{R}_{q}\mathbf{Y}_{u}^{\dagger
})^{JM}(\mathbf{Y}_{d}\mathbf{R}_{q}\mathbf{Y}_{u}^{\dagger})^{KN}\smallskip$
& \multicolumn{1}{c}{$\det(g_{U}^{\dagger})$}\\
\multicolumn{1}{c}{$%
%TCIMACRO{\TeXButton{lambda}{\bm{\lambda}}}%
%BeginExpansion
\bm{\lambda}%
%EndExpansion
_{3}^{\prime\prime IJK}$\rule{0in}{0.17in}} & $\varepsilon^{LMN}%
(\mathbf{Y}_{u}\mathbf{R}_{q})^{IL}(\mathbf{Y}_{d}\mathbf{R}_{q}%
)^{JM}(\mathbf{Y}_{d}\mathbf{R}_{q})^{KN}\smallskip$ &
\multicolumn{1}{c}{$\det(g_{Q}^{\dagger})$}\\
\multicolumn{1}{c}{$%
%TCIMACRO{\TeXButton{lambda}{\bm{\lambda}}}%
%BeginExpansion
\bm{\lambda}%
%EndExpansion
_{4}^{\prime\prime IJK}$\rule{0in}{0.17in}} & $\varepsilon^{LMN}%
\varepsilon^{ABC}\varepsilon^{DEF}(\mathbf{R}_{q}\mathbf{Y}_{d}^{\dagger
})^{LD}(\mathbf{R}_{q}\mathbf{Y}_{u}^{\dagger})^{MA}(\mathbf{R}_{q}%
\mathbf{Y}_{u}^{\dagger})^{NB}(\mathbf{R}_{u})^{IC}(\mathbf{R}_{d}%
)^{JE}(\mathbf{R}_{d})^{KF}\smallskip$ & \multicolumn{1}{c}{$\det(g_{Q}%
g_{U}^{\dagger}g_{D}^{\dagger})$}\\
\multicolumn{1}{c}{$%
%TCIMACRO{\TeXButton{lambda}{\bm{\lambda}}}%
%BeginExpansion
\bm{\lambda}%
%EndExpansion
_{5}^{\prime\prime IJK}$\rule{0in}{0.17in}} & $\varepsilon^{STU}%
\varepsilon^{ABC}\varepsilon^{DEF}\mathbf{\epsilon}^{\prime LMNPQR}%
(\mathbf{R}_{q}\mathbf{Y}_{u}^{\dagger})^{LS}(\mathbf{R}_{q}\mathbf{Y}%
_{u}^{\dagger})^{PT}(\mathbf{R}_{u})^{IU}$ & \\
\multicolumn{1}{c}{} & $\;\;\;\;\;\;\times(\mathbf{R}_{q}\mathbf{Y}%
_{d}^{\dagger})^{MA}(\mathbf{R}_{q}\mathbf{Y}_{d}^{\dagger})^{QB}%
(\mathbf{R}_{d})^{JC}(\mathbf{R}_{q}\mathbf{Y}_{d}^{\dagger})^{ND}%
(\mathbf{R}_{q}\mathbf{Y}_{d}^{\dagger})^{RE}(\mathbf{R}_{d})^{KF}\smallskip$
& $\det(g_{Q}^{2}g_{U}^{\dagger}g_{D}^{\dagger2})$\\\hline
\end{tabular}
\ $%
%TCIMACRO{\TeXButton{caption}{\caption
%{Superpotential $RPV$ terms under the MFV hypothesis. For
%$\bm{\lambda}_{i}^{IJK}$, it is understood that contributions must be
%antisymmetrized under $I\leftrightarrow J$ while, similarly, $\bm{\lambda}
%_{i}^{\prime\prime IJK}{}$ must be antisymmetrized under $J\leftrightarrow K$.
%The explicit MFV expansion for each structure is obtained by summing over the
%possible insertions of the $\mathbf{R}_{i}$ terms of
%Eqs.(\ref{Xq}), (\ref{Xl}) and (\ref{Rright}%
%), and inserting arbitrary $\mathcal{O}(1)$ MFV
%coefficients in front of each term of this sum. Finally, the $\varepsilon
%$-tensors each act in one of the five $SU(3)$, hence break a specific $U(1)$,
%as indicated in the last column.}}}%
%BeginExpansion
\caption{Superpotential $RPV$ terms under the MFV hypothesis. For
$\bm{\lambda}_{i}^{IJK}$, it is understood that contributions must be
antisymmetrized under $I\leftrightarrow J$ while, similarly, $\bm{\lambda}
_{i}^{\prime\prime IJK}{}$ must be antisymmetrized under $J\leftrightarrow K$.
The explicit MFV expansion for each structure is obtained by summing over the
possible insertions of the $\mathbf{R}_{i}$ terms of
Eqs.(\ref{Xq}), (\ref{Xl}) and (\ref{Rright}%
), and inserting arbitrary $\mathcal{O}(1)$ MFV
coefficients in front of each term of this sum. Finally, the $\varepsilon
$-tensors each act in one of the five $SU(3)$, hence break a specific $U(1)$,
as indicated in the last column.}%
%EndExpansion
\label{TableRPV}%
%TCIMACRO{\TeXButton{E}{\end{table}}}%
%BeginExpansion
\end{table}%
%EndExpansion

\subsubsection{Application to RPV couplings}

The MFV expansions for the $RPV$ terms of the superpotential are collected in
Table \ref{TableRPV}, where the intermediate spurion $\mathbf{\bar{\Upsilon}%
}_{\nu}$, transforming as $\mathbf{\bar{\Upsilon}}_{\nu}\rightarrow\det
(g_{L})\mathbf{\bar{\Upsilon}}_{\nu}g_{L}^{\dagger}$, is%
\begin{equation}
\mathbf{\bar{\Upsilon}}_{\nu.}^{I}\equiv\varepsilon^{QMJ}\mathbf{\Upsilon
}_{\nu}^{\dagger PN}(\mathbf{R}_{\ell})^{QN}(\mathbf{R}_{\ell})^{MP}%
(\mathbf{R}_{\ell})^{JI}=\varepsilon^{QMJ}(\mathbf{R}_{\ell}\mathbf{\Upsilon
}_{\nu}^{\dagger}\mathbf{R}_{\ell}^{T})^{MQ}(\mathbf{R}_{\ell})^{JI}\;,
\label{InterSpur}%
\end{equation}
and the $\mathbf{\epsilon}$ and $\epsilon^{\prime}$ tensors,%
\begin{align}
\mathbf{\epsilon}^{ABCDEF}  &  \equiv\varepsilon^{ACE}\varepsilon
^{BDF},\;\varepsilon^{EFD}\varepsilon^{ABC},\;\varepsilon^{EFA}\varepsilon
^{BCD}\;,\label{TensorT}\\
\mathbf{\epsilon}^{\prime LMNPQR}  &  \equiv\varepsilon^{LMN}\varepsilon
^{PQR},\;\varepsilon^{LMQ}\varepsilon^{NPR},\;\varepsilon^{LMP}\varepsilon
^{NQR}\;,
\end{align}
stands for the three inequivalent contractions in the $SU(3)_{L}$ and
$SU(3)_{Q}$-space, respectively. For $\mathbf{\epsilon}$, there is a fourth
possible contraction, $\varepsilon^{ABE}\varepsilon^{CDF}$, which gives back $%
%TCIMACRO{\TeXButton{lambda}{\bm{\lambda}}}%
%BeginExpansion
\bm{\lambda}%
%EndExpansion
_{3}$. Structures involving more $\varepsilon$-tensors can be reduced to those
of Table \ref{TableRPV} using Eq.(\ref{epsi}).

We do not write down explicitly the MFV expansion for the $RPV$ soft-breaking
terms since they can be readily obtained from Table \ref{TableRPV}:
$\mathbf{b}^{\prime}$ and $\mathbf{m}_{Ld}^{2}$ transform as $%
%TCIMACRO{\TeXButton{mu}{\bm{\mu}}}%
%BeginExpansion
\bm{\mu}%
%EndExpansion
^{\prime}$, while $\mathbf{A}$, $\mathbf{A}^{\prime}$, and $\mathbf{A}%
^{\prime\prime}$ transform as $%
%TCIMACRO{\TeXButton{lambda}{\bm{\lambda}}}%
%BeginExpansion
\bm{\lambda}%
%EndExpansion
$, $%
%TCIMACRO{\TeXButton{lambda}{\bm{\lambda}}}%
%BeginExpansion
\bm{\lambda}%
%EndExpansion
^{\prime}$, and $%
%TCIMACRO{\TeXButton{lambda}{\bm{\lambda}}}%
%BeginExpansion
\bm{\lambda}%
%EndExpansion
^{\prime\prime}$, respectively. The normalization of the dimensionful
couplings will be addressed in Sec.\ref{Scales}. For now, we just state that $%
%TCIMACRO{\TeXButton{mu}{\bm{\mu}}}%
%BeginExpansion
\bm{\mu}%
%EndExpansion
^{\prime}$, $\mathbf{b}^{\prime}$ and $\mathbf{m}_{Ld}^{2}$ are normalized
with respect to $\mu$, $b$ and $m_{H_{d}}^{2}$, respectively, while trilinear
soft-breaking terms are all normalized by the supersymmetry-breaking scale
$A_{0}$ (see Eq.(\ref{RPCsoft})). The MFV coefficients are then dimensionless
and assumed to be of $\mathcal{O}(1)$.

Reminiscent of the fact that $RPV$ operators break either baryon or lepton
number, each of them involves at least one $\varepsilon$-tensor (the
$\mathbf{R}_{i}$ are neutral under all $U(1)$'s). However, in terms of the
$U(1)$'s acting on the individual fields, we have the freedom to decide in
which direction to break $B$ and $L$. As shown in Table \ref{TableRPV}, the
$\Delta L=1$ structures break $U(1)_{L}$ or both $U(1)_{E}$ and $U(1)_{L}$,
while $\Delta B=1$ structures break $U(1)_{U}$, $U(1)_{D}$ and/or $U(1)_{Q}$.

We are not forced to simultaneously break all these $U(1)$'s, only one per
sector is needed. In particular, we can require the $U(1)$ for all the
right-handed fields to remain exact. Alternatively, we can choose to maintain
only $U(1)_{D}$ and $U(1)_{E}$, which are intimately connected with
$U(1)_{PQ}$. Indeed, if we do not assign a $U(1)_{PQ}$ charge to $H_{d}$, it
is then born by the Yukawas $\mathbf{Y}_{d}$ and $\mathbf{Y}_{\ell}$. Terms
which violate $U(1)_{D}$ or $U(1)_{E}$ are then precisely those which violate
$U(1)_{PQ}$.

In practice, enforcing one of the $U(1)$'s amounts to suppress the structures
which break it by powers of $\det(\mathbf{Y}_{u})$, $\det(\mathbf{Y}_{d})$ or
$\det(\mathbf{Y}_{\ell})$. These determinants always involve the light-fermion
masses, and are very small even at large $\tan\beta$. For example, enforcing
$U(1)_{U,D,E}$ suppresses all $\varepsilon$-structures in the $RPC$
soft-breaking terms of Eq.(\ref{RPCsoft}), while it leaves only $%
%TCIMACRO{\TeXButton{mu}{\bm{\mu}}}%
%BeginExpansion
\bm{\mu}%
%EndExpansion
_{1}^{\prime}$, $%
%TCIMACRO{\TeXButton{lambda}{\bm{\lambda}}}%
%BeginExpansion
\bm{\lambda}%
%EndExpansion
_{1,2}$, $%
%TCIMACRO{\TeXButton{lambda}{\bm{\lambda}}}%
%BeginExpansion
\bm{\lambda}%
%EndExpansion
_{1}^{\prime}$ and $%
%TCIMACRO{\TeXButton{lambda}{\bm{\lambda}}}%
%BeginExpansion
\bm{\lambda}%
%EndExpansion
_{3}^{\prime\prime}$ as dominant $RPV$ structures. Finally, if we decide to
enforce $U(1)_{L}$, all $\Delta L=1$ structures get suppressed by at least one
power of $\det(\mathbf{Y}_{\ell})$. This global suppression of the $\Delta
L=1$ sector is possible because $\mathbf{\Upsilon}_{\nu}$ is transforming
non-trivially only under $SU(3)_{L}$.

Due to the antisymmetry of $\varepsilon$-tensors, some of the terms vanish
identically, so that the bases in Table \ref{TableRPV} are not fully reduced
algebraically. Nevertheless, the number of terms for each $RPV$ couplings is
much larger than their true degrees of freedom. However, in many cases, only a
handful of operators are dominant and need to be kept, as we will see in the
next section. For now, we just note that if $\tan\beta$ is not large, one can
neglect $\mathbf{Y}_{d}$ compared to $\mathbf{Y}_{u}$, while if $M_{R}$ is
smaller than about $10^{13}$ GeV, $\mathbf{Y}_{\nu}^{\dagger}\mathbf{Y}_{\nu}$
is negligible and can be dropped everywhere. One then remains with about $10$
complex parameters to describe $%
%TCIMACRO{\TeXButton{mu}{\bm{\mu}}}%
%BeginExpansion
\bm{\mu}%
%EndExpansion
^{\prime}$, $%
%TCIMACRO{\TeXButton{lambda}{\bm{\lambda}}}%
%BeginExpansion
\bm{\lambda}%
%EndExpansion
$, $%
%TCIMACRO{\TeXButton{lambda}{\bm{\lambda}}}%
%BeginExpansion
\bm{\lambda}%
%EndExpansion
^{\prime}$, and $%
%TCIMACRO{\TeXButton{lambda}{\bm{\lambda}}}%
%BeginExpansion
\bm{\lambda}%
%EndExpansion
^{\prime\prime}$ couplings (see Appendix A), depending on which of the
$U(1)$'s remain exact.

\subsection{Natural scales for the RPV-MFV coefficients\label{Scales}}

It is remarkable that all the $RPV$ couplings can be generated out of the
minimal set of spurions needed to account for the known fermion masses and
mixings. In addition, it appears that there is a fundamental distinction
between the baryon and lepton number violating terms. Indeed, $\Delta L=1$
couplings are strictly forbidden as long as $m_{\nu}=0$, since the $\nu_{L}$
Majorana mass, transforming as $(\bar{6},1)$, is definitely needed to get
invariants under $G_{\ell}$(\footnote{This is not the only possible choice of
spurions in the lepton sector. In Ref.\cite{DavidsonP06}, the Majorana mass
arises from $RPV$ couplings, promoted to spurions, while there is no need to
extend the flavor group at high-energy. However, that approach is not so
predictive for LFV effects, and further, cannot explain proton stability. In
the present work, the smallness of $RPV$ effects originates from their MFV
structures, and does not have to be imposed separately.}). Then, the seesaw
mechanism not only suppresses neutrino masses, it suppresses all $\Delta
L\neq0$ effects. On the other hand, $\Delta B=1$ couplings can be readily
parametrized in terms of the usual quark Yukawas.

Naturalness demands all MFV coefficients to be of $\mathcal{O}(1)$, but leaves
open the overall normalization of dimensionful couplings like $%
%TCIMACRO{\TeXButton{mu}{\bm{\mu}}}%
%BeginExpansion
\bm{\mu}%
%EndExpansion
^{\prime}$ or $RPV$ soft-breaking terms. In the present section, this issue
will be analyzed from several perspectives, showing that the naive
normalization of dimensionful $RPV$ couplings in terms of their $RPC$
counterparts is the most natural. In this respect, it is worth to mention that
RGE invariance cannot help much. Indeed, the MFV expansions are stable under
the RGE, and, further, the $\Delta B=1$ and $\Delta L=1$ sectors are
decoupled\cite{MartinV93}.

\subsubsection{Basis independence, sneutrino VEV's and neutrino masses}

When lepton-number is not conserved, the left-handed lepton doublet $L^{I}$
and the Higgs doublet $H_{d}$ have the same quantum numbers and can mix. A
priori, the Lagrangian fields do not correspond to the physical Higgs and
lepton states. In other words, defining the four-component vector
$\phi^{\alpha}=(H_{d},L^{I})$, the physics is invariant if we carry out the
field redefinition\cite{HallS84}%
\begin{equation}
\phi^{\alpha}\rightarrow U_{\beta}^{\alpha}\phi^{\beta}\;,
\end{equation}
with $U\in SU(4)$. Obviously, the gauge sector is invariant, but what we call
the $RPC$ and ($\Delta L=1$) $RPV$ sectors get mixed. Indeed, one can
immediately see from Eqs.(\ref{WRPC}) and (\ref{WRPV}) that a change of basis
modifies the relative size of $\mathbf{Y}_{\ell}$ and $%
%TCIMACRO{\TeXButton{lambda}{\bm{\lambda}}}%
%BeginExpansion
\bm{\lambda}%
%EndExpansion
$, $\mathbf{Y}_{d}$ and $%
%TCIMACRO{\TeXButton{lambda}{\bm{\lambda}}}%
%BeginExpansion
\bm{\lambda}%
%EndExpansion
^{\prime}$, and $\mu$ and $%
%TCIMACRO{\TeXButton{mu}{\bm{\mu}}}%
%BeginExpansion
\bm{\mu}%
%EndExpansion
^{\prime}$ (the soft-breaking terms are similarly affected). Since we can, for
example, choose a basis in which one bilinear term is rotated away, there are
too many parameters and only some combinations of them are physical. We will
now check that the expansions obtained in the previous section satisfy the MFV
principle despite these ambiguities.

If the $RPV$ couplings take, in some basis, the MFV forms obtained in the
previous section, the sneutrino VEV's $\langle\nu^{I}\rangle$ are in general
non-vanishing. We must thus check that rotating these VEV's away only amounts
to $\mathcal{O}(1)$ redefinitions of the MFV coefficients for all $RPV$
couplings. Let us consider for now only the $RPC$ and $RPV$ bilinear terms,
Eqs.(\ref{RPCbi}) and Eq.(\ref{RPVbi}), written in four-component notation%
\begin{equation}
W\ni\bar{\mu}_{\alpha}\left(  H_{u}\phi^{\alpha}\right)  ,\;\;\mathcal{L}%
_{\text{soft}}\ni-(\bar{b}_{\alpha}(H_{u}\phi^{\alpha})+h.c.)-\bar{m}%
_{\alpha\beta}^{2}\phi^{\alpha\dagger}\phi^{\beta}\;,
\end{equation}
where%
\begin{equation}
\bar{\mu}^{\alpha}=(\mu,%
%TCIMACRO{\TeXButton{mu}{\bm{\mu}}}%
%BeginExpansion
\bm{\mu}%
%EndExpansion
^{\prime}),\;\bar{b}^{\alpha}=(b,\mathbf{b}^{\prime}),\;\bar{m}_{\alpha\beta
}^{2}=\left(
\begin{array}
[c]{cc}%
m_{H_{d}}^{2} & \mathbf{m}_{Ld}^{2}\\
(\mathbf{m}_{Ld}^{2})^{\dagger} & \mathbf{m}_{L}^{2}%
\end{array}
\right)  \;.
\end{equation}
In Ref.\cite{BanksGNN95}, it was shown that if $\bar{b}\ $is proportional to
$\bar{\mu}$, and $\bar{\mu}$ is an eigenvector of $\bar{m}^{2}$, then we can
choose a basis in which the vacuum expectation values $\langle\phi^{\alpha
}\rangle$ are aligned with $\bar{\mu}^{\alpha}$, in particular we can choose
$\bar{\mu}^{\alpha}=(\mu,0,0,0)$, $\bar{b}^{\alpha}=\left(  b,0,0,0\right)  $
and $\langle\phi^{\alpha}\rangle=(v_{d},0,0,0)$. Now, if the MFV expansions
for $%
%TCIMACRO{\TeXButton{mu}{\bm{\mu}}}%
%BeginExpansion
\bm{\mu}%
%EndExpansion
^{\prime}$, $\mathbf{b}^{\prime}$ and $\mathbf{m}_{Ld}^{2}$ are normalized
with respect to $\mu$, $b$ and $m_{H_{d}}^{2}$, respectively, we are in a
situation of near-alignment and sneutrino VEV's are very small:%
\begin{equation}
\bar{\mu}^{\alpha}=(\mu,\mu\mathbf{\bar{\Upsilon}}_{\nu}),\;\bar{b}^{\alpha
}=(b,b\mathbf{\bar{\Upsilon}}_{\nu}^{\prime}),\;\mathbf{m}_{Ld}^{2}=m_{H_{d}%
}^{2}\mathbf{\bar{\Upsilon}}_{\nu}^{\prime\prime}\;\rightarrow\bar{\mu}%
\times\bar{b}=\mathcal{O}(\mathbf{\bar{\Upsilon}}_{\nu}),\;\bar{m}%
_{\alpha\beta}^{2}\bar{\mu}^{\beta}=m_{H_{d}}^{2}\bar{\mu}_{\alpha
}+\mathcal{O}(\mathbf{\bar{\Upsilon}}_{\nu})\,,
\end{equation}
and therefore $\langle\nu^{I}\rangle\sim\mathcal{O}(v_{d}\mathbf{\bar
{\Upsilon}}_{\nu}^{I})$. In other words, the misalignment is entirely due to
the $\mathcal{O}(1)$ differences between the MFV coefficients of the
$\mathbf{\bar{\Upsilon}}_{\nu}$ structures of $\bar{\mu}$, $\bar{b}$ and
$\bar{m}^{2}$. To rotate away these VEV's, consider the change of basis%
\begin{equation}
U=\left(
\begin{array}
[c]{cc}%
1 & -\varepsilon^{I}\\
\varepsilon^{\ast I} & 1_{3\times3}%
\end{array}
\right)  \;,
\end{equation}
with $\varepsilon^{I}=\langle\nu^{I}\rangle/v_{d}=a\mathbf{\bar{\Upsilon}%
}_{\nu}^{I}$. The constant $a$ is of $\mathcal{O}(1)$ and we set it to one for
simplicity. The impact for all $RPC$ terms is completely negligible, while the
redefined $RPV$ terms automatically satisfy their MFV expansions:%
\begin{gather}
\delta%
%TCIMACRO{\TeXButton{mu}{\bm{\mu}}}%
%BeginExpansion
\bm{\mu}%
%EndExpansion
^{\prime I}=\mu\mathbf{\bar{\Upsilon}}_{\nu}^{I},\;\delta%
%TCIMACRO{\TeXButton{lambda}{\bm{\lambda}}}%
%BeginExpansion
\bm{\lambda}%
%EndExpansion
^{IJK}=\mathbf{\bar{\Upsilon}}_{\nu}^{I}(\mathbf{Y}_{\ell})^{KJ}%
-(I\overset{\,\,}{\leftrightarrow}J),\;\delta%
%TCIMACRO{\TeXButton{lambda}{\bm{\lambda}}}%
%BeginExpansion
\bm{\lambda}%
%EndExpansion
^{\prime IJK}=\mathbf{\bar{\Upsilon}}_{\nu}^{I}(\mathbf{Y}_{d})^{KJ}\;,\\
\delta\mathbf{b}^{\prime I}=b\mathbf{\bar{\Upsilon}}_{\nu}^{I},\;\delta
\mathbf{m}_{Ld}^{2}=m_{H_{d}}^{2}\mathbf{\bar{\Upsilon}}_{\nu},\;\delta
\mathbf{A}^{IJK}=\mathbf{\bar{\Upsilon}}_{\nu}^{I}(\mathbf{A}_{\ell}%
)^{KJ}-(I\overset{\,\,}{\leftrightarrow}J),\;\delta\mathbf{A}^{\prime
IJK}=\mathbf{\bar{\Upsilon}}_{\nu}^{I}(\mathbf{A}_{d})^{KJ}\;.
\end{gather}
This rotation thus only induces $\mathcal{O}(1)$ shifts in the values of the
MFV coefficients of $RPV$ terms, provided the $RPV$ soft-breaking trilinear
terms are normalized by $A_{0}$. Note that by the same reasoning, one can also
see that the MFV expansion is stable if one of the bilinear terms is rotated away.

Given the freedom to rotate the $RPC$ and $RPV$ couplings, it could happen
that the MFV structure is hidden. In other words, $RPV$ couplings could be
very large but fine-tuned with $RPC$ terms, such that moving to the
$\langle\nu^{I}\rangle=0$ basis, they would again assume their MFV forms. This
latter form is more natural in the sense that the $\Delta L=1$ couplings are
then of the same order of magnitude as the physical, basis-independent
parameters describing $\Delta L=1$ effects\cite{DavidsonE96}. Indeed, for
example, the two basis-independent angles $\xi$ and $\zeta$ tuning the lepton
-- Higgsino and slepton -- Higgs mixings are both $\mathcal{O}(\mathbf{\bar
{\Upsilon}}_{\nu})$\cite{BanksGNN95,GrossmanH98}:
\begin{equation}
\cos\xi=\frac{1}{\left|  \bar{\mu}\right|  v_{d}}\sum_{\alpha}\mu_{\alpha
}v^{\alpha}\rightarrow\sin\xi=\mathcal{O}(\mathbf{\bar{\Upsilon}}_{\nu
}),\;\;\cos\zeta=\frac{1}{\left|  \bar{b}\right|  v_{d}}\sum_{\alpha}%
b_{\alpha}v^{\alpha}\rightarrow\sin\zeta=\mathcal{O}(\mathbf{\bar{\Upsilon}%
}_{\nu})\;.
\end{equation}

Finally, since these angles are very small, the impact of $RPV$ couplings on
charged lepton or neutrino masses is negligible, and the background values for
the spurions can be fixed as in Eq.(\ref{Background}). This is obvious for the
charged leptons, while for the $\Delta L=2$ neutrino masses, it is necessarily
quadratic in $\Delta L=1$ effects, i.e. $\mathcal{O}(\mathbf{\Upsilon}_{\nu
}^{2})$. For example, the tree-level mixing induced by the $RPV$ bilinear
terms scales as $\tan^{2}\xi$\cite{BanksGNN95}, while those generated at the
loop-level by the $RPV$ trilinear terms scale as $%
%TCIMACRO{\TeXButton{lambda}{\bm{\lambda}}}%
%BeginExpansion
\bm{\lambda}%
%EndExpansion
^{2}$ or $%
%TCIMACRO{\TeXButton{lambda}{\bm{\lambda}}}%
%BeginExpansion
\bm{\lambda}%
%EndExpansion
^{\prime2}$.

\subsubsection{High-energy scales and higher-dimensional
operators\label{SectionScales}}

In the present work, the $(\bar{6},1)$ spurion is normalized as
$\mathbf{\Upsilon}_{\nu}=U^{\ast}\mathbf{m}_{\nu}U^{\dagger}/v_{u}$, so that
it lies on the same footing as the other fermion masses, see
Eq.(\ref{Background}). Consequently, all $\Delta L=1$ couplings are very
suppressed if MFV holds, since at least one power of $\mathbf{\Upsilon}_{\nu}$
is needed to make them invariant under $G_{f}$. This is the most natural and
model-independent assumption because, as said previously, the MSSM spurion
content does not allow for $\Delta L=1$ couplings in the $m_{\nu}=0$ limit.

At the same time though, the seesaw mechanism is responsible for the smallness
of $m_{\nu}$, and call for additional degrees of freedom at the scale
$\Lambda_{\Delta L=2}\sim M_{R}$. Therefore, it is tempting to associate this
scale also with $\Delta L=1$ effects, i.e. to imagine that they arise from
some non-trivial dynamics at the $M_{R}$ scale. To concoct such a model is not
trivial and lies beyond the purpose of the present article. However, it should
be clear that if $\Lambda_{\Delta L=1}\sim\Lambda_{\Delta L=2}$, the same
spurion $\mathbf{\Upsilon}_{\nu}$ can be used to parametrize both $\Delta L=1$
and $\Delta L=2$ couplings.

Alternatively, if $\Lambda_{\Delta L=2}>>\Lambda_{\Delta L=1}$, large
compensating factors would arise, since the spurion to be used is then
$r\times\mathbf{\Upsilon}_{\nu}$ with $r=\Lambda_{\Delta L=2}/\Lambda_{\Delta
L=1}$. Note that this corresponds to the rescaling%
\begin{equation}
\bar{\mu}^{\alpha}=\left(  \mu,r\mu\mathbf{\bar{\Upsilon}}_{\nu}\right)
,\,\bar{b}^{\alpha}=\left(  b,rb\mathbf{\bar{\Upsilon}}_{\nu}\right)
,\,\mathbf{m}_{Ld}^{2}=rm_{H_{d}}^{2}\mathbf{\bar{\Upsilon}}_{\nu},\,%
%TCIMACRO{\TeXButton{lambda}{\bm{\lambda}}}%
%BeginExpansion
\bm{\lambda}%
%EndExpansion
,%
%TCIMACRO{\TeXButton{lambda}{\bm{\lambda}}}%
%BeginExpansion
\bm{\lambda}%
%EndExpansion
^{\prime}\sim r\mathcal{O}(\mathbf{\Upsilon}_{\nu}),\,\mathbf{A}%
,\mathbf{A}^{\prime}\sim rA_{0}\mathcal{O}(\mathbf{\Upsilon}_{\nu})\;,
\end{equation}
which is compatible with the MFV expansion (the developments of the previous
section remain essentially unchanged, with now $\langle\nu^{I}\rangle
=\mathcal{O}(rv_{d}\mathbf{\bar{\Upsilon}}_{\nu}^{I})$). Of course, $r$ should
not be too large, otherwise $\Delta L=1$ couplings would contribute to the
neutrino mass, thereby invalidating Eq.(\ref{Background}).

In the present work, we stick to the minimal hypothesis and assume
$\Lambda_{\Delta L=1}\sim\Lambda_{\Delta L=2}$. Ultimately, it is the
comparison with experimental constraints which will tell us if this is viable,
or will give us clues as to the scale at which $\Delta L=1$ effects arise, and
hopefully about the dynamics going on there.

In the $\Delta B=1$ sector, there is no seesaw mechanism at play and therefore
no clue as to the mechanism behind their generation. In the present work, we
treat $\Delta B=1$ couplings on the same footing as $RPC$ terms, i.e. we
accept that baryon number is simply not conserved. However, one should keep in
mind that all $\Delta B=1$ MFV coefficients could very well be suppressed or
enhanced by some ratio of scales, or suppressed by some gauge couplings,
$a_{i}\sim\mathcal{O}(g^{2}/4\pi)$.

If we imagine that there is a non-trivial lepton-number violating dynamics
going on at the high-energy scale, it is natural to expect that integrating
out the heavy degrees of freedom leads to additional dimension-five
operators\cite{IbanezR91}. Let us concentrate on the $RPC$ dimension-five
operators in the superpotential%
\begin{equation}
W_{\dim-5}\ni\frac{%
%TCIMACRO{\TeXButton{kappa}{\bm{\kappa}}}%
%BeginExpansion
\bm{\kappa}%
%EndExpansion
_{1}^{IJKL}}{\Lambda_{\Delta L=1}}(Q^{I}Q^{J})(Q^{K}L^{L})+\frac{%
%TCIMACRO{\TeXButton{kappa}{\bm{\kappa}}}%
%BeginExpansion
\bm{\kappa}%
%EndExpansion
_{2}^{IJKL}}{\Lambda_{\Delta L=1}}(D^{I}U^{J}U^{K})E^{L}+\frac{%
%TCIMACRO{\TeXButton{kappa}{\bm{\kappa}}}%
%BeginExpansion
\bm{\kappa}%
%EndExpansion
_{5}^{IJ}}{\Lambda_{\Delta L=2}}(L^{I}H_{u})(L^{J}H_{u})\;.
\end{equation}
The operator $%
%TCIMACRO{\TeXButton{kappa}{\bm{\kappa}}}%
%BeginExpansion
\bm{\kappa}%
%EndExpansion
_{5}$ corresponds to the one arising from the integration of the right-handed
neutrinos, Eq.(\ref{Dim5nu}), with the scale $\Lambda_{\Delta L=2}$ then given
by $M_{R}$.

The overall scale of $%
%TCIMACRO{\TeXButton{kappa}{\bm{\kappa}}}%
%BeginExpansion
\bm{\kappa}%
%EndExpansion
_{1}$ and $%
%TCIMACRO{\TeXButton{kappa}{\bm{\kappa}}}%
%BeginExpansion
\bm{\kappa}%
%EndExpansion
_{2}$ is simply $\Lambda_{\Delta L=1}$ even though they are also breaking
baryon number, since we do not associate any particular scale to $\Delta B=1$
effects. If we assume again that $\Lambda_{\Delta L=1}\sim\Lambda_{\Delta
L=2}$, the operators $%
%TCIMACRO{\TeXButton{kappa}{\bm{\kappa}}}%
%BeginExpansion
\bm{\kappa}%
%EndExpansion
_{1}$ and $%
%TCIMACRO{\TeXButton{kappa}{\bm{\kappa}}}%
%BeginExpansion
\bm{\kappa}%
%EndExpansion
_{2}$ could induce proton decay at an unacceptable rate. However, we think
that if enforcing MFV is sufficient to separately suppress $\Delta L=1$ and
$\Delta B=1$ interactions so as to pass experimental bounds on proton decay,
the same should be true for $%
%TCIMACRO{\TeXButton{kappa}{\bm{\kappa}}}%
%BeginExpansion
\bm{\kappa}%
%EndExpansion
_{1}$ and $%
%TCIMACRO{\TeXButton{kappa}{\bm{\kappa}}}%
%BeginExpansion
\bm{\kappa}%
%EndExpansion
_{2}$. Indeed, the flavor group $G_{f}$ factorizes as $G_{q}\times G_{\ell
}\times G_{1}$, hence it makes no difference whether MFV is used to
parametrize a product of $\Delta L=1$ and $\Delta B=1$ operators, or a single
$\Delta L=1$, $\Delta B=1$ operator. Explicitly, the MFV expansions are%
\begin{align}
\frac{%
%TCIMACRO{\TeXButton{kappa}{\bm{\kappa}}}%
%BeginExpansion
\bm{\kappa}%
%EndExpansion
_{1}^{IJKL}}{\Lambda_{\Delta L=1}}  &  =\frac{1}{v_{u}}\varepsilon
^{MNP}(\mathbf{R}_{q})^{MI}(\mathbf{R}_{q})^{NJ}(\mathbf{R}_{q})^{PK}%
\mathbf{\bar{\Upsilon}}_{\nu}^{L}\;,\\
\frac{%
%TCIMACRO{\TeXButton{kappa}{\bm{\kappa}}}%
%BeginExpansion
\bm{\kappa}%
%EndExpansion
_{2}^{IJKL}}{\Lambda_{\Delta L=1}}  &  =\frac{1}{v_{u}}(%
%TCIMACRO{\TeXButton{lambda}{\bm{\lambda}}}%
%BeginExpansion
\bm{\lambda}%
%EndExpansion
^{\prime\prime IJK})_{u\leftrightarrow d}(\mathbf{Y}_{\ell}\mathbf{\bar
{\Upsilon}}_{\nu}^{\dagger})^{L}\;,
\end{align}
where $(%
%TCIMACRO{\TeXButton{lambda}{\bm{\lambda}}}%
%BeginExpansion
\bm{\lambda}%
%EndExpansion
^{\prime\prime IJK})_{u\leftrightarrow d}$ is obtained from Table
\ref{TableRPV} by interchanging $\mathbf{Y}_{d}\leftrightarrow\mathbf{Y}_{u}$
and $\mathbf{R}_{d}\leftrightarrow\mathbf{R}_{u}$. Obviously, their structures
are very similar to simple products of $\Delta L=1$ and $\Delta B=1$
couplings. To check that $%
%TCIMACRO{\TeXButton{kappa}{\bm{\kappa}}}%
%BeginExpansion
\bm{\kappa}%
%EndExpansion
_{1}$ and $%
%TCIMACRO{\TeXButton{kappa}{\bm{\kappa}}}%
%BeginExpansion
\bm{\kappa}%
%EndExpansion
_{2}$ pass the experimental bounds if $%
%TCIMACRO{\TeXButton{lambda}{\bm{\lambda}}}%
%BeginExpansion
\bm{\lambda}%
%EndExpansion
$, $%
%TCIMACRO{\TeXButton{lambda}{\bm{\lambda}}}%
%BeginExpansion
\bm{\lambda}%
%EndExpansion
^{\prime}$ and $%
%TCIMACRO{\TeXButton{lambda}{\bm{\lambda}}}%
%BeginExpansion
\bm{\lambda}%
%EndExpansion
^{\prime\prime}$ do would require a detailed analysis, which lies out of our
main purpose. Indeed, contrary to the renormalizable $RPV$ couplings, the $%
%TCIMACRO{\TeXButton{kappa}{\bm{\kappa}}}%
%BeginExpansion
\bm{\kappa}%
%EndExpansion
_{1}$ and $%
%TCIMACRO{\TeXButton{kappa}{\bm{\kappa}}}%
%BeginExpansion
\bm{\kappa}%
%EndExpansion
_{2}$ contributions to proton decay arise only at the loop-level, since these
interactions preserve R-parity. They thus depend also on the parameters of the
MSSM gauge sector (see Ref.\cite{IbanezR91} for a discussion). Further, the
suppressions due to flavor-mixing are not necessarily to be found in the $%
%TCIMACRO{\TeXButton{kappa}{\bm{\kappa}}}%
%BeginExpansion
\bm{\kappa}%
%EndExpansion
_{i}$ themselves. For instance, the $%
%TCIMACRO{\TeXButton{kappa}{\bm{\kappa}}}%
%BeginExpansion
\bm{\kappa}%
%EndExpansion
_{1}^{123L}$ coupling is not suppressed by CKM or light-quark mass factors,
but necessarily involve a third-generation (s)quark, hence its contribution to
proton decay will nevertheless involve some flavor-mixings.

Because of these flavor-mixings and loop-suppression factors, the
contributions to proton decay from non-renormalizable operators should be
subleading compared to the $%
%TCIMACRO{\TeXButton{lambda}{\bm{\lambda}}}%
%BeginExpansion
\bm{\lambda}%
%EndExpansion
\times%
%TCIMACRO{\TeXButton{lambda}{\bm{\lambda}}}%
%BeginExpansion
\bm{\lambda}%
%EndExpansion
^{\prime\prime}$ and $%
%TCIMACRO{\TeXButton{lambda}{\bm{\lambda}}}%
%BeginExpansion
\bm{\lambda}%
%EndExpansion
^{\prime}\times%
%TCIMACRO{\TeXButton{lambda}{\bm{\lambda}}}%
%BeginExpansion
\bm{\lambda}%
%EndExpansion
^{\prime\prime}$ tree-level contributions, and will not be considered anymore
here. However, it is remarkable that the MFV principle can offer a common
solution to both problems, irrespective of whether the couplings are $RPV$ or
$RPC$. As said, this is essentially because the flavor group factorizes as
$G_{q}\times G_{\ell}$, while R-parity acts additively for a given coupling.

\section{Phenomenological consequences for the R-parity violating MSSM}

When applied to $RPV$ couplings, the MFV hypothesis necessarily makes use of
$\varepsilon$-tensors to contract the spurions to $G_{q}\times G_{\ell}$
singlets. Because of its antisymmetry, the couplings are then in general
proportional to products of fermion masses of more than one generation. Given
the strong hierarchy among these masses, MFV tends to strongly suppress all
$RPV$ couplings. It is the purpose of the present section to analyze under
which circumstances this suppression is sufficient to pass experimental bounds
on $\Delta B=1$ and $\Delta L=1$ processes, especially proton decay. Once
established, the impact of these conditions on the possible $RPV$ effects at
colliders, and the connections with FCNC or LFV effects will be briefly commented.

It should be remarked also that naive expectations like $\lambda^{\prime\prime
IJK}\sim O(m_{u}^{I}m_{d}^{J}m_{d}^{K}/v_{u}v_{d}^{2})$ are not valid in MFV,
because the generation indices are twisted by the antisymmetric $\varepsilon
$-tensors. MFV thus implies a very peculiar form of helicity-suppression with,
for example, a coupling involving the up-quark tuned by $m_{t}/v_{u}$. Also,
MFV predicts that the $\Delta L=1$ couplings are all proportional to products
of neutrino and charged lepton masses.

\subsection{Experimental information for the spurions}

The flavor symmetry permits to rotate the spurions to their background values,
Eq.(\ref{Background}), which can be fixed in terms of experimental values. For
the Yukawa $\mathbf{Y}_{u,d,\ell}$, we take the quark and charged lepton
masses, as well as the CKM matrix elements from Ref.\cite{PDG}. For the
neutrino spurions $\mathbf{Y}_{\nu}^{\dagger}\mathbf{Y}_{\nu}$ and
$\mathbf{\Upsilon}_{\nu}$, we start from the neutrino mixing parameters of the
best-fit of Ref.\cite{GonzalezM07}%
\begin{align}
\Delta m_{21}^{2}  &  =\Delta m_{\odot}^{2}=7.9_{-0.28}^{+0.27}\times
10^{-5}\,\text{eV}^{2},\;|\Delta m_{31}^{2}|=\Delta m_{atm}^{2}=(2.6\pm
0.2)\times10^{-3}\,\text{eV}^{2}\;,\nonumber\\
\theta_{12}  &  =\theta_{\odot}=\left(  33.7\pm1.3\right)
%TCIMACRO{\U{b0}}%
%BeginExpansion
{{}^\circ}%
%EndExpansion
,\;\theta_{23}=\theta_{atm}=\left(  43.3_{-3.8}^{+4.3}\right)
%TCIMACRO{\U{b0}}%
%BeginExpansion
{{}^\circ}%
%EndExpansion
,\;\theta_{13}=\left(  0_{-0}^{+5.2}\right)
%TCIMACRO{\U{b0}}%
%BeginExpansion
{{}^\circ}%
%EndExpansion
\;.
\end{align}
In a first approximation, since we are only interested in the order of
magnitude of the $RPV$ couplings, we can neglect the small $\theta_{13}$ as
well as the $CP$-violating phase, and fix the atmospheric angle at
$\theta_{atm}=45%
%TCIMACRO{\U{b0}}%
%BeginExpansion
{{}^\circ}%
%EndExpansion
$ (maximal mixing), such that the $PMNS$ mixing matrix takes the simple form:%
\begin{equation}
U\simeq\left(
\begin{array}
[c]{ccc}%
c_{\odot} & s_{\odot} & 0\\
-s_{\odot}/\sqrt{2} & c_{\odot}/\sqrt{2} & 1/\sqrt{2}\\
s_{\odot}/\sqrt{2} & -c_{\odot}/\sqrt{2} & 1/\sqrt{2}%
\end{array}
\right)  \;,
\end{equation}
where $s_{\odot}=\sin\theta_{\odot}$, $c_{\odot}=\cos\theta_{\odot}$ and
$\tan\theta_{\odot}\simeq2/3$. Under these approximations,%
\begin{equation}
\mathbf{\Upsilon}_{\nu}=\frac{1}{v_{u}}\left(  m_{\nu}\mathbf{1}+\frac{\Delta
m_{21}}{\sqrt{2}\left(  1+t_{\odot}^{2}\right)  }\left(
\begin{array}
[c]{ccc}%
\sqrt{2}t_{\odot}^{2} & t_{\odot} & -t_{\odot}\\
t_{\odot} & 1/\sqrt{2} & -1/\sqrt{2}\\
-t_{\odot} & -1/\sqrt{2} & 1/\sqrt{2}%
\end{array}
\right)  +\frac{\Delta m_{31}}{2}\left(
\begin{array}
[c]{ccc}%
0 & 0 & 0\\
0 & 1 & 1\\
0 & 1 & 1
\end{array}
\right)  \right)  \;, \label{SpurionNu}%
\end{equation}
where $m_{\nu1}=m_{\nu}$, $m_{\nu2}=m_{\nu}+\Delta m_{21}$, $m_{\nu3}=m_{\nu
}+\Delta m_{31}$. The neutrino mass-scale $m_{\nu}$ is unknown, but should not
exceed about $1$ eV if the cosmological bound $\sum_{i}m_{i}\lesssim1\;$eV
holds\cite{Cosmo}. Depending on the spectrum (i.e., whether $\nu_{1}$ or
$\nu_{3}$ is the lightest neutrino), the mass-differences $\Delta m_{21}$ and
$\Delta m_{31}$ are related to the mixing parameters as%
\begin{equation}
\Delta m_{21}=(\Delta m_{\odot}^{2}+m_{\nu}^{2})^{1/2}-m_{\nu}>0,\;\left\{
\begin{array}
[c]{l}%
\Delta m_{31}=(\Delta m_{atm}^{2}+m_{\nu}^{2})^{1/2}-m_{\nu}%
>0\;\;\text{(Normal),}\\
\Delta m_{31}=(m_{\nu}^{2}-\Delta m_{atm}^{2})^{1/2}-m_{\nu}%
<0\;\;\text{(Inverted).}%
\end{array}
\right.  \label{SpurionNu2}%
\end{equation}
Therefore, for fixed $\Delta m_{\odot}^{2}$ and $\Delta m_{atm}^{2}$,
off-diagonal elements of $\mathbf{\Upsilon}_{\nu}$ quickly decrease with
increasing $m_{\nu}$, with the maximum being%
\begin{align}
m_{\nu_{1}}  &  =0\;\text{(Normal)}:\Delta m_{21}=(\Delta m_{\odot}^{2}%
)^{1/2},\;\Delta m_{31}=(\Delta m_{atm}^{2})^{1/2}\;\text{\ },\;\\
m_{\nu_{3}}  &  =0\;\text{(Inverted)}:\Delta m_{21}=\frac{\Delta m_{\odot}%
^{2}}{2(\Delta m_{atm}^{2})^{1/2}},\Delta m_{31}=-(\Delta m_{atm}^{2}%
)^{1/2}\;.
\end{align}
For the normal spectrum, $m_{\nu}$ varies between $0$ and about $1$ eV, while
in the inverted hierarchy, it varies between $(\Delta m_{atm}^{2})^{1/2}$ and
about $1$ eV. Therefore, the off-diagonal elements of $\mathbf{\Upsilon}_{\nu
}$ are typically smaller for the inverted spectrum, and we will not consider
that case anymore.

Contrary to the other spurions, $\mathbf{Y}_{\nu}^{\dagger}\mathbf{Y}_{\nu}$
cannot be fixed entirely in terms of experimentally known quantities, since
neutrino masses and mixings only give us access to $\mathbf{Y}_{\nu}%
^{T}\mathbf{M}^{-1}\mathbf{Y}_{\nu}$. Following
Refs.\cite{PascoliPY03,CiriglianoIP06}, the overall ambiguity in
$\mathbf{Y}_{\nu}^{\dagger}\mathbf{Y}_{\nu}$ can be parametrized in terms of
three phases:%
\begin{equation}
\mathbf{Y}_{\nu}^{\dagger}\mathbf{Y}_{\nu}=\frac{M_{R}}{v_{u}^{2}%
}U\,\mathbf{m}_{\nu}^{1/2}\,\mathbf{H}^{2}\,\mathbf{m}_{\nu}^{1/2}%
\,U^{\dagger}\;,\;\;\mathbf{H}=e^{i\mathbf{\Phi}},\;\;\mathbf{\Phi}=\left(
\begin{array}
[c]{ccc}%
0 & \phi_{1} & \phi_{2}\\
-\phi_{1} & 0 & \phi_{3}\\
-\phi_{2} & -\phi_{3} & 0
\end{array}
\right)  \;.
\end{equation}
We work in the $CP$-limit, $\phi_{i}=0$, such that $\mathbf{Y}_{\nu}^{\dagger
}\mathbf{Y}_{\nu}\overset{CP}{=}M_{R}\mathbf{\Upsilon}_{\nu}/v_{u}$, i.e. both
neutrino spurions are real, parallel and symmetric. Corrections induced by the
$CP$-phases $\phi_{i}$ are assumed to be small\footnote{Expanding
$\mathbf{H}=\mathbf{1}+i\mathbf{\Phi}$, these phases can be introduced through
a new spurion $\Delta_{\mathbf{\Phi}}=M_{R}U\,\mathbf{m}_{\nu}^{1/2}%
\,\mathbf{\Phi}\,\mathbf{m}_{\nu}^{1/2}\,U^{\dagger}/v_{u}^{2}$, still
suppressed by the neutrino masses, and presumably smaller than $M_{R}%
\mathbf{\Upsilon}_{\nu}/v_{u}$ if the phases are not large. For our leading
order analysis, the perturbations induced by $\mathbf{\Delta}_{\mathbf{\Phi}}$
are neglected.}, while those due to $\delta_{13}$ are always suppressed by
$\sin\theta_{13}$. In practice, taking $\mathbf{Y}_{\nu}^{\dagger}%
\mathbf{Y}_{\nu}$ and $\mathbf{\Upsilon}_{\nu}$ aligned greatly reduces the
number of structures, since, for example%
\begin{equation}
\varepsilon^{IJK}(\mathbf{Y}_{\nu}^{\dagger}\mathbf{Y}_{\nu}\mathbf{\Upsilon
}_{\nu}^{\dagger})^{IJ}\overset{CP}{=}0\;, \label{Symm}%
\end{equation}
by symmetry. Also, the Cayley-Hamilton relation Eq.(\ref{CH}) can be used to
discard products of two or more $\mathbf{Y}_{\nu}^{\dagger}\mathbf{Y}_{\nu
}\mathbf{\ }$with $\mathbf{\Upsilon}_{\nu}$. Further, the $m_{\nu}\mathbf{1}$
piece of $\mathbf{Y}_{\nu}^{\dagger}\mathbf{Y}_{\nu}$ can be dropped since the
identity is already part of $\mathbf{R}_{\ell}$. Finally, for our perturbative
expansion to make sense, we must ensure that $\mathbf{Y}_{\nu}^{\dagger
}\mathbf{Y}_{\nu}\lesssim1$, which translates as%
\begin{equation}
\dfrac{\max[m_{\nu},|\Delta m_{31}|]}{1\text{\thinspace eV}}\frac{M_{R}%
}{10^{13}\,\text{GeV}}\lesssim3\;. \label{UB}%
\end{equation}
For $m_{\nu}\simeq0$, $M_{R}$ can be at most $\sim5\times10^{14}$, while for
$M_{R}\lesssim10^{13}$, the spurion $\mathbf{Y}_{\nu}^{\dagger}\mathbf{Y}%
_{\nu}$ is very suppressed as all its elements are tuned by $\Delta m_{21}$
and $\Delta m_{31}$.

\subsection{The reduced MFV expansion and order-of-magnitude estimates}

The first step to get the order of magnitude of the couplings is to reduce the
MFV operator bases constructed in the previous section. Indeed, they all
involve operators which are very suppressed once experimental values for the
spurions are plugged in. Specifically, an operator can be neglected if it
induces only small corrections to the entries of all possible linear
combinations of the others, within a tolerance of about $5\%-10\%$.

However, this reduction is not trivial because the operators to include, as
well as the order of magnitude of the $RPV$ couplings, crucially depend on
$\tan\beta$, $M_{R}$ and $m_{\nu}$. Generically, the number of operators
increases with increasing $\tan\beta$, $M_{R}$, or decreasing $m_{\nu}$, and
having several non-aligned spurions makes the $RPV$ couplings less
hierarchical. Because of these dependences, to give a general and
simultaneously minimal basis is not possible. For example, an operator can be
dominant if $\tan\beta$ increases, but become subleading if $m_{\nu}$
decreases. Further, when $\tan\beta$, $M_{R}$ are large and $m_{\nu}$ is
small, the number of independent and dominant operators is in general
comparable to the number of free parameters needed to fully specify the $RPV$ couplings.

For these reasons, we prefer to analyze four extreme situations numerically%
\begin{equation}%
\begin{array}
[c]{rlll}%
\text{Case I}: & \tan\beta=5, & M_{R}=10^{12}\;\text{GeV}, & m_{\nu
}=0.5\;\text{eV,}\\
\text{Case II}: & \tan\beta=50, & M_{R}=10^{12}\;\text{GeV}, & m_{\nu
}=0.5\;\text{eV,}\\
\text{Case III}: & \tan\beta=5, & M_{R}=2\times10^{14}\;\text{GeV}, & m_{\nu
}=0\;\text{eV,}\\
\text{Case IV}: & \tan\beta=50,\; & M_{R}=2\times10^{14}\;\text{GeV},\; &
m_{\nu}=0\;\text{eV.}%
\end{array}
\label{Cases}%
\end{equation}
Case II and IV maximize the impact of $\mathbf{Y}_{d}$ and $\mathbf{Y}_{\ell}%
$, while Case III\ and IV maximize that of $\mathbf{Y}_{\nu}^{\dagger
}\mathbf{Y}_{\nu}$ and $\mathbf{\Upsilon}_{\nu}$. Note that since we are only
interested in order-of-magnitude estimates, we neglect the non-holomorphic
corrections to the Yukawas induced at large $\tan\beta$\cite{HallRS93}, and
keep the background values fixed as in Eq.(\ref{Background}).

In the Appendix A, we construct analytically the minimal basis for
\begin{equation}
\text{Case V}:\tan\beta\lesssim20,\;M_{R}\lesssim2\times10^{13}\;\text{GeV}%
,\;m_{\nu}\gtrsim0.05\;\text{eV\ .}%
\end{equation}
In that region of parameter space, $\mathbf{Y}_{\nu}^{\dagger}\mathbf{Y}_{\nu
}$ is always subleading and the basis is rather simple.

\begin{table}[t]
\centering                                         $%
\begin{tabular}
[c]{cccccc}\hline
\rule{0in}{0.17in} & $\mathbf{\bar{\Upsilon}}_{\nu}^{I}$ &
$\boldsymbol{\lambda}  _{1}^{IJK}$ & $\boldsymbol{\lambda}  _{2}^{IJK}$ &
$\boldsymbol{\lambda}  _{3}^{IJK}$ & $\boldsymbol{\lambda}  _{4}^{IJK}%
$\\\hline
$%
\begin{array}
[c]{c}%
\text{Scaling}\\
\text{in }\tan\beta
\end{array}
$ & $\tan^{2}\beta$ & $\tan^{3}\beta$ & $\tan\beta$ & $\tan^{4}\beta$ &
$\tan^{2}\beta$\\\hline
$\text{Case I}$ & ${\small \left(
\begin{array}
[c]{c}%
17\\
19\\
21
\end{array}
\right)  }$ & ${\small \left(
\begin{array}
[c]{ccc}%
23 & 19 & 23\\
30 & 24 & 20\\
26 & 24 & 18
\end{array}
\right)  }$ & ${\small \left(
\begin{array}
[c]{ccc}%
21 & 17 & 13\\
16 & 18 & 17\\
21 & 14 & 15
\end{array}
\right)  }$ & ${\small \left(
\begin{array}
[c]{ccc}%
22 & 23 & 28\\
28 & 27 & 26\\
25 & 27 & 24
\end{array}
\right)  }$ & ${\small \left(
\begin{array}
[c]{ccc}%
19 & 20 & 19\\
15 & 22 & 23\\
19 & 17 & 21
\end{array}
\right)  }$\\
Case II & ${\small \left(
\begin{array}
[c]{c}%
15\\
17\\
19
\end{array}
\right)  }$ & ${\small \left(
\begin{array}
[c]{ccc}%
20 & 16 & 20\\
27 & 21 & 17\\
23 & 21 & 15
\end{array}
\right)  }$ & ${\small \left(
\begin{array}
[c]{ccc}%
20 & 16 & 12\\
15 & 17 & 16\\
20 & 13 & 14
\end{array}
\right)  }$ & ${\small \left(
\begin{array}
[c]{ccc}%
18 & 19 & 24\\
24 & 23 & 22\\
21 & 23 & 20
\end{array}
\right)  }$ & ${\small \left(
\begin{array}
[c]{ccc}%
17 & 18 & 17\\
13 & 20 & 21\\
17 & 15 & 19
\end{array}
\right)  }$\\
Case III & ${\small \left(
\begin{array}
[c]{c}%
16\\
17\\
18
\end{array}
\right)  }$ & ${\small \left(
\begin{array}
[c]{ccc}%
21 & 18 & 18\\
23 & 20 & 18\\
22 & 19 & 17
\end{array}
\right)  }$ & ${\small \left(
\begin{array}
[c]{ccc}%
19 & 15 & 14\\
19 & 16 & 15\\
19 & 15 & 14
\end{array}
\right)  }$ & ${\small \left(
\begin{array}
[c]{ccc}%
20 & 21 & 24\\
22 & 23 & 24\\
21 & 22 & 23
\end{array}
\right)  }$ & ${\small \left(
\begin{array}
[c]{ccc}%
17 & 19 & 20\\
17 & 20 & 21\\
17 & 19 & 20
\end{array}
\right)  }$\\
Case IV & ${\small \left(
\begin{array}
[c]{c}%
14\\
15\\
16
\end{array}
\right)  }$ & ${\small \left(
\begin{array}
[c]{ccc}%
18 & 15 & 15\\
20 & 17 & 15\\
19 & 16 & 14
\end{array}
\right)  }$ & ${\small \left(
\begin{array}
[c]{ccc}%
18 & 14 & 13\\
18 & 15 & 14\\
18 & 14 & 13
\end{array}
\right)  }$ & ${\small \left(
\begin{array}
[c]{ccc}%
16 & 18 & 20\\
18 & 19 & 20\\
17 & 18 & 19
\end{array}
\right)  }$ & ${\small \left(
\begin{array}
[c]{ccc}%
15 & 17 & 18\\
15 & 18 & 19\\
15 & 17 & 18
\end{array}
\right)  }$\\\hline
\end{tabular}
\ \ \ $\caption{The order of magnitude of the intermediate spurion
$\mathbf{\bar{\Upsilon}}_{\nu}$, Eq.(\ref{InterSpur}), and of the
$\boldsymbol{\lambda}  ^{IJK}(L^{I}L^{J})E^{K}$ couplings, separately for each
$\varepsilon$-structure of Table \ref{TableRPV}. Entries are to be understood
as $x\equiv\mathcal{O}(10^{-x})$. The matrix entries correspond to the $J,K$
indices, with $I$ fixed as $I=2,3,1$ for $J=1,2,3$, respectively.}%
\label{TRPV1}%
\end{table}

In Tables \ref{TRPV1}--\ref{TRPV3}, the order of magnitude of the couplings
are indicated, separately for each $\varepsilon$-structure. These values
correspond to the contribution of the numerically dominant operator for each
individual coupling. In other words, after expanding the terms of each
structure of Table \ref{TableRPV}, plugging in the numerical values for the
spurions, we find for each value of the $I,J,K$ indices a numerical polynomial
in the MFV coefficients. Assuming these coefficients are of $\mathcal{O}(1)$,
and barring large interferences among operators, we take the largest term of
this polynomial as the order of magnitude of the coupling. The notation in
Table \ref{TRPV1} for $%
%TCIMACRO{\TeXButton{lambda}{\bm{\lambda}}}%
%BeginExpansion
\bm{\lambda}%
%EndExpansion
^{IJK}$, which is antisymmetric under $I\leftrightarrow J$, is to arrange its
nine degrees of freedom in a $3\times3$ matrix $%
%TCIMACRO{\TeXButton{lambda}{\bm{\lambda}}}%
%BeginExpansion
\bm{\lambda}%
%EndExpansion
^{IJK}=\mathbf{X}^{JK}$, with $I=2,3,1$ for $J=1,2,3$, respectively.
Similarly, $%
%TCIMACRO{\TeXButton{lambda}{\bm{\lambda}}}%
%BeginExpansion
\bm{\lambda}%
%EndExpansion
^{\prime\prime IJK}$ is antisymmetric under $J\leftrightarrow K$ and written
in Table \ref{TRPV3} as $%
%TCIMACRO{\TeXButton{lambda}{\bm{\lambda}}}%
%BeginExpansion
\bm{\lambda}%
%EndExpansion
^{\prime\prime IJK}=\mathbf{X}^{IJ}$ with $K=2,3,1$ for $J=1,2,3$,
respectively. Note that this coupling is insensitive to $M_{R}$ and $m_{\nu}$,
only $\tan\beta$ is relevant.

It is immediately apparent from these tables that MFV is quite powerful to
predict the overall scales in terms of $\tan\beta$, $M_{R}$ and $m_{\nu}$, as
well as the hierarchies within each structure. This is because the
$\varepsilon$-tensors twist the hierarchy of the $\mathbf{Y}_{u,d,\ell}$
spurions in a specific way. Of course, for terms involving many different
structures like $%
%TCIMACRO{\TeXButton{lambda}{\bm{\lambda}}}%
%BeginExpansion
\bm{\lambda}%
%EndExpansion
$ and $%
%TCIMACRO{\TeXButton{lambda}{\bm{\lambda}}}%
%BeginExpansion
\bm{\lambda}%
%EndExpansion
^{\prime\prime}$, these hierarchies are somewhat blurred. However, if we
enforce anyone of the $U(1)$'s, some of the structures get suppressed by
additional $\det(\mathbf{Y}_{u,d,\ell})$ factors, which are very small in all
cases:
\begin{subequations}
\label{Det}%
\begin{align}
\tan\beta\overset{}{=}5  &  :\det(\mathbf{Y}_{u})\simeq10^{-7},\;\det
(\mathbf{Y}_{d})\simeq10^{-7},\;\det(\mathbf{Y}_{\ell})\simeq10^{-9}\;,\\
\tan\beta\overset{}{=}50  &  :\det(\mathbf{Y}_{u})\simeq10^{-7},\;\det
(\mathbf{Y}_{d})\simeq10^{-4},\;\det(\mathbf{Y}_{\ell})\simeq10^{-6}\;.
\end{align}
This can have important consequences for phenomenology, as we will see in the
next subsection.

\begin{table}[t]
\centering              $%
\begin{tabular}
[c]{ccc}\hline
\rule{0in}{0.17in} & $\boldsymbol{\lambda}  _{1}^{\prime IJK}$ &
$\boldsymbol{\lambda}  _{2}^{\prime IJK}$\\\hline
$%
\begin{array}
[c]{c}%
\text{Scaling}\\
\text{in }\tan\beta
\end{array}
$ & $\tan^{3}\beta$ & $\tan^{4}\beta$\\\hline
Case I/III & $\mathbf{\bar{\Upsilon}}_{\nu}^{I}\times{\small \left(
\begin{array}
[c]{ccc}%
4 & 6 & 3\\
7 & 3 & 3\\
6 & 4 & 1
\end{array}
\right)  }^{JK}$ & $\mathbf{\bar{\Upsilon}}_{\nu}^{I}\times{\small \left(
\begin{array}
[c]{ccc}%
3 & 8 & 8\\
7 & 5 & 7\\
5 & 6 & 6
\end{array}
\right)  }^{JK\dfrac{{}}{{}}}$\\
Case II/IV & $\mathbf{\bar{\Upsilon}}_{\nu}^{I}\times{\small \left(
\begin{array}
[c]{ccc}%
3 & 5 & 2\\
6 & 2 & 1\\
5 & 3 & 0
\end{array}
\right)  }^{JK}$ & $\mathbf{\bar{\Upsilon}}_{\nu}^{I}\times{\small \left(
\begin{array}
[c]{ccc}%
1 & 6 & 6\\
5 & 3 & 5\\
3 & 4 & 4
\end{array}
\right)  }^{JK}$\\\hline
\end{tabular}
\ $\caption{The order of magnitude of the $\boldsymbol{\lambda}  ^{\prime
IJK}(L^{I}Q^{J})D^{K}$ couplings, for each $\varepsilon$-structure of Table
\ref{TableRPV}. Entries are to be understood as $x\equiv\mathcal{O}(10^{-x})$.
The spurion $\mathbf{\bar{\Upsilon}}_{\nu}$ for each case is given in Table
\ref{TRPV1}.}%
\end{table}

In Tables \ref{TRPV1}--\ref{TRPV3}, we also give the dominant $\tan\beta$
behavior, i.e. the behavior of the simplest term for each structure (i.e., the
skeleton). Some individual elements may scale with higher powers of $\tan
\beta$ than indicated if they are sensitive to the presence of $\mathbf{Y}%
_{\ell}$ or $\mathbf{Y}_{d}$. Anyway, MFV predicts that all $RPV$ couplings
scale at least linearly with $\tan\beta$. Indeed, in addition to the explicit
powers of $\mathbf{Y}_{d,\ell}$ shown in Table \ref{TableRPV}, Eq.(\ref{Symm})
implies the presence of at least one power of $\mathbf{Y_{\ell}^{\dagger}%
Y}_{\ell}$ in $\mathbf{\bar{\Upsilon}}_{\nu}$. On the other hand, as
$\tan\beta$ increases, the hierarchies within each structure are not much affected.

To maximize the impact of $\mathbf{Y}_{\nu}^{\dagger}\mathbf{Y}_{\nu}$ and
$\mathbf{\Upsilon}_{\nu}$ we set $m_{\nu}=0$. This may seem counter-intuitive
given Eq.(\ref{SpurionNu}). However, we already noticed that the $m_{\nu
}\mathbf{1}$ piece is irrelevant for $\mathbf{Y}_{\nu}^{\dagger}%
\mathbf{Y}_{\nu}$ because the identity is already part of $\mathbf{R}_{\ell}$.
Concerning $\mathbf{\Upsilon}_{\nu}$, in most cases it occurs through
$\mathbf{\bar{\Upsilon}}_{\nu}$, whose dominant term is%
\end{subequations}
\begin{equation}
\mathbf{\bar{\Upsilon}}_{\nu}^{I}=\varepsilon^{QMI}(\mathbf{Y}_{\ell}%
^{\dagger}\mathbf{Y}_{\ell}\mathbf{\Upsilon}_{\nu}^{\dagger})^{MQ}+\,...
\end{equation}
Therefore, the $m_{\nu}\mathbf{1}$ piece of $\mathbf{\Upsilon}_{\nu}$ again
disappears since $\mathbf{Y}_{\ell}^{\dagger}\mathbf{Y}_{\ell}$ is symmetric.
Then, apart$\ $from $%
%TCIMACRO{\TeXButton{lambda}{\bm{\lambda}}}%
%BeginExpansion
\bm{\lambda}%
%EndExpansion
_{2,4}^{213,321,132}$(\footnote{This can be seen by looking at the skeleton
for these structures, and setting $\Upsilon_{\nu}=m_{\nu}\mathbf{1}$. Then,
$\lambda_{2}^{IJK}=\varepsilon^{IJN}(\mathbf{Y}_{\ell})^{KN}$ and $\lambda
_{4}^{IJK}=\varepsilon^{KLM}(\mathbf{Y}_{\ell}^{\dagger})^{JL}(\mathbf{Y}%
_{\ell}^{\dagger})^{IM}$, which are non-zero only for $IJK,JIK=213,321,132$%
.}), the $\Delta L=1$ structures are all tuned essentially by $\Delta m_{21}$
and $\Delta m_{31}$, and these mass-differences are maximized when $m_{\nu}%
=0$. In that case, the hierarchies are softened, because the neutrino spurions
have large non-diagonal elements (see Eq.(\ref{SpurionNu})). Finally, we take
$M_{R}$ large enough to make $\mathbf{Y}_{\nu}^{\dagger}\mathbf{Y}_{\nu}$
competitive. However, apart from softening the hierarchies, the presence of
the $\mathbf{Y}_{\nu}^{\dagger}\mathbf{Y}_{\nu}$ spurion has no impact on the
order of magnitude of the dominant entries.

Let us look in more details at each coupling. The bilinear terms are all
proportional to $\mathbf{\bar{\Upsilon}}_{\nu}$, Eq.(\ref{InterSpur}), which
is smaller than naively expected from its proportionality to neutrino masses
because of the presence of $\mathbf{Y}_{\ell}^{\dagger}\mathbf{Y}_{\ell}$.
Note that the hierarchy of $\mathbf{Y}_{\ell}^{\dagger}\mathbf{Y}_{\ell}$ gets
inverted by the $\varepsilon$-tensor, resulting in the order of magnitudes
shown in the first column of Table \ref{TRPV1}.

\begin{table}[t]
\centering                                      $%
\begin{tabular}
[c]{ccccc}\hline
\rule{0in}{0.17in} & $\boldsymbol{\lambda}  _{1}^{\prime\prime IJK}$ &
$\boldsymbol{\lambda}  _{2}^{\prime\prime IJK}$ & $\boldsymbol{\lambda}
_{3}^{\prime\prime IJK}$ & $\boldsymbol{\lambda}  _{4}^{\prime\prime IJK}%
$\\\hline
$%
\begin{array}
[c]{c}%
\text{Scaling}\\
\text{in }\tan\beta
\end{array}
$ & $\tan\beta$ & $\tan^{2}\beta$ & $\tan^{2}\beta$ & $\tan\beta$\\\hline
Case I/III & $\left(  {\small
\begin{array}
[c]{ccc}%
8 & 8 & 8\\
4 & 6 & 5\\
1 & 6 & 4
\end{array}
}\right)  $ & $\left(  {\small
\begin{array}
[c]{ccc}%
11 & 6 & 7\\
12 & 9 & 9\\
13 & 12 & 13
\end{array}
}\right)  $ & $\left(  {\small
\begin{array}
[c]{ccc}%
13 & 8 & 10\\
10 & 6 & 7\\
6 & 5 & 6
\end{array}
}\right)  $ & $\left(  {\small
\begin{array}
[c]{ccc}%
5 & 5 & 5\\
7 & 9 & 7\\
7 & 12 & 10
\end{array}
}\right)  $\\
Case II/IV & $\left(  {\small
\begin{array}
[c]{ccc}%
7 & 7 & 7\\
3 & 5 & 4\\
0 & 5 & 3
\end{array}
}\right)  $ & $\left(  {\small
\begin{array}
[c]{ccc}%
9 & 4 & 5\\
10 & 7 & 7\\
11 & 10 & 11
\end{array}
}\right)  $ & $\left(  {\small
\begin{array}
[c]{ccc}%
11 & 6 & 8\\
8 & 4 & 5\\
4 & 3 & 4
\end{array}
}\right)  $ & $\left(  {\small
\begin{array}
[c]{ccc}%
4 & 4 & 4\\
6 & 8 & 6\\
6 & 11 & 9
\end{array}
}\right)  $\\\hline
\end{tabular}
\ \ \ $\caption{The order of magnitude of the $\boldsymbol{\lambda}
^{\prime\prime IJK}U^{I}D^{J}D^{K}$ couplings, for each $\varepsilon
$-structure of Table \ref{TableRPV}. Entries are to be understood as
$x\equiv\mathcal{O}(10^{-x})$. The matrix entries correspond to the $I,J$
indices, with $K$ fixed as $K=2,3,1$ for $J=1,2,3$, respectively.}%
\label{TRPV3}%
\end{table}

For $%
%TCIMACRO{\TeXButton{lambda}{\bm{\lambda}}}%
%BeginExpansion
\bm{\lambda}%
%EndExpansion
$, the first structure, proportional to $\mathbf{\bar{\Upsilon}}_{\nu}$, is in
general smaller than the others. Then, if we impose $U(1)_{E}$, the third and
fourth structures are suppressed by an additional $\det(\mathbf{Y}_{\ell})$
factor and can be dropped. In that case, only $%
%TCIMACRO{\TeXButton{lambda}{\bm{\lambda}}}%
%BeginExpansion
\bm{\lambda}%
%EndExpansion
_{2}$ needs to be kept, with its very definite hierarchy.

The $%
%TCIMACRO{\TeXButton{lambda}{\bm{\lambda}}}%
%BeginExpansion
\bm{\lambda}%
%EndExpansion
^{\prime}$ coupling is simply obtained as the direct product $%
%TCIMACRO{\TeXButton{lambda}{\bm{\lambda}}}%
%BeginExpansion
\bm{\lambda}%
%EndExpansion
^{\prime IJK}=\mathbf{\bar{\Upsilon}}_{\nu}^{I}\left(  \mathbf{A}_{d}\right)
^{KJ}$. Therefore, beside the suppression brought in by $\mathbf{\bar
{\Upsilon}}_{\nu}$, it is further suppressed by down-quark masses. In Table 3,
we only write explicitly $\left(  \mathbf{A}_{d}\right)  ^{KJ}$ for $\tan
\beta=5$ and $\tan\beta=50$. It is interesting to remark that the second
structure, corresponding to the $\varepsilon$-terms for $\mathbf{A}_{d}$ in
Eq.(\ref{RPCsoft}), has an inverted hierarchy compared to the first (see
Appendix B for the numerical analysis of the $RPC$ trilinear terms including
the $\varepsilon$-structures).

Finally, the $\Delta B=1$ couplings, given in Table \ref{TRPV3}, are clearly
much larger than those breaking lepton-number. Nevertheless, a priori, one
would have expected them to be all of $\mathcal{O}(1)$, so MFV suppresses them
significantly. Only in the first structure, which involves the least number of
Yukawas, there is an entry of $\mathcal{O}(1)$ when it involves the (s)top, $%
%TCIMACRO{\TeXButton{lambda}{\bm{\lambda}}}%
%BeginExpansion
\bm{\lambda}%
%EndExpansion
_{312}^{\prime\prime}$. If we impose the $U(1)_{D}$ symmetry, $\Delta B=1$
couplings are all much smaller, $%
%TCIMACRO{\TeXButton{lambda}{\bm{\lambda}}}%
%BeginExpansion
\bm{\lambda}%
%EndExpansion
^{\prime\prime}\lesssim10^{-5}\,(10^{-3})$ for $\tan\beta=5\,(50)$,
respectively. The complicated structure $%
%TCIMACRO{\TeXButton{lambda}{\bm{\lambda}}}%
%BeginExpansion
\bm{\lambda}%
%EndExpansion
_{5}^{\prime\prime}$ is smaller than $%
%TCIMACRO{\TeXButton{lambda}{\bm{\lambda}}}%
%BeginExpansion
\bm{\lambda}%
%EndExpansion
_{4}^{\prime\prime}$ by at least three orders of magnitude, even at large
$\tan\beta$, and can be safely neglected.

\subsection{Bounds on $\Delta B=1$ nucleon decays and consequences}

The strongest experimental bounds come from the non-observation of $\Delta
B=1$ nucleon decay, setting constraints on certain combinations of the $\Delta
L=1$ and $\Delta B=1$ couplings, and from $\Delta B=2$ neutron--antineutron
oscillations, directly constraining the $%
%TCIMACRO{\TeXButton{lambda}{\bm{\lambda}}}%
%BeginExpansion
\bm{\lambda}%
%EndExpansion
^{\prime\prime}$ couplings. For their part, taken alone, the $\Delta L=1$
couplings easily pass all experimental constraints since they are suppressed
by the small neutrino masses. In the present section, we set all $RPV$
soft-breaking terms to zero, and we assume that the $\Delta B=1$ nucleon
decays are only into quark and lepton final states.

\begin{table}[t]
\centering
\begin{tabular}
[c]{rlcccccccccccc}\hline
\multicolumn{2}{c}{Approximate\ bound\ \ \ $\;\;\;\;\;\;\;$} &
\multicolumn{3}{c}{Case I} & \multicolumn{3}{c}{Case II} &
\multicolumn{3}{c}{Case III} & \multicolumn{3}{c}{Case IV}\\
\multicolumn{2}{c}{} & A & B & C & A & B & C & A & B & C & A & B & C\\\hline
$\left|  \boldsymbol{\lambda}  _{J1I}^{\prime}\boldsymbol{\lambda}
_{11I}^{\prime\prime\ast}\right|  $ & $\lesssim10^{-27}\tilde{d}_{R,I}^{2}$ &
\multicolumn{1}{|c}{$24$} & $27$ & \multicolumn{1}{l}{$30$} &
\multicolumn{1}{|c}{$\mathbf{20}$} & $\mathbf{22}$ & $24$ &
\multicolumn{1}{|c}{$\mathbf{23}$} & $26$ & $28$ &
\multicolumn{1}{|c}{$\mathbf{19}$} & $\mathbf{21}$ & $\mathbf{23}$\\
$\left|  \boldsymbol{\lambda}  _{J2I}^{\prime}\boldsymbol{\lambda}
_{11I}^{\prime\prime\ast}\right|  $ & $\lesssim10^{-27}\tilde{d}_{R,I}^{2}$ &
\multicolumn{1}{|c}{$24$} & $26$ & \multicolumn{1}{l}{$29$} &
\multicolumn{1}{|c}{$\mathbf{20}$} & $\mathbf{21}$ & $24$ &
\multicolumn{1}{|c}{$\mathbf{22}$} & $25$ & $28$ &
\multicolumn{1}{|c}{$\mathbf{18}$} & $\mathbf{20}$ & $\mathbf{23}$\\
$\left|  \boldsymbol{\lambda}  _{M1I}^{\prime}\boldsymbol{\lambda}
_{12I}^{\prime\prime\ast}\right|  $ & $\lesssim10^{-27}\tilde{d}_{R,I}^{2}$ &
\multicolumn{1}{|c}{$25$} & $25$ & \multicolumn{1}{l}{$28$} &
\multicolumn{1}{|c}{$\mathbf{20}$} & $\mathbf{20}$ & $\mathbf{23}$ &
\multicolumn{1}{|c}{$\mathbf{23}$} & $24$ & $27$ &
\multicolumn{1}{|c}{$\mathbf{18}$} & $\mathbf{19}$ & $\mathbf{22}$\\
$\left|  \boldsymbol{\lambda}  _{IJ1}^{\prime}\boldsymbol{\lambda}
_{11J}^{\prime\prime\ast}\right|  $ & $\lesssim10^{-27}\tilde{d}_{L,J}%
^{2}(\delta_{J}^{D})^{-1}$ & \multicolumn{1}{|c}{$27$} & $30$ &
\multicolumn{1}{l}{$32$} & \multicolumn{1}{|c}{$\mathbf{22}$} & $25$ & $27$ &
\multicolumn{1}{|c}{$25$} & $28$ & $31$ & \multicolumn{1}{|c}{$\mathbf{20}$} &
$\mathbf{23}$ & $26$\\
$\left|  \boldsymbol{\lambda}  _{IJ2}^{\prime}\boldsymbol{\lambda}
_{11J}^{\prime\prime\ast}\right|  $ & $\lesssim10^{-27}\tilde{d}_{L,J}%
^{2}(\delta_{J}^{D})^{-1}$ & \multicolumn{1}{|c}{$24$} & $28$ &
\multicolumn{1}{l}{$31$} & \multicolumn{1}{|c}{$\mathbf{20}$} & $\mathbf{23}$
& $25$ & \multicolumn{1}{|c}{$23$} & $27$ & $29$ &
\multicolumn{1}{|c}{$\mathbf{19}$} & $\mathbf{21}$ & $24$\\
$\left|  \boldsymbol{\lambda}  _{I31}^{\prime}\boldsymbol{\lambda}
_{123}^{\prime\prime\ast}\right|  $ & $\lesssim10^{-27}\tilde{b}_{L}%
^{2}(\delta_{3}^{D})^{-1}$ & \multicolumn{1}{|c}{$27$} & $28$ &
\multicolumn{1}{l}{$31$} & \multicolumn{1}{|c}{$\mathbf{21}$} & $\mathbf{23}$
& $26$ & \multicolumn{1}{|c}{$26$} & $27$ & $29$ &
\multicolumn{1}{|c}{$\mathbf{20}$} & $\mathbf{22}$ & $24$\\
\multicolumn{1}{l}{$\left|  \boldsymbol{\lambda}  _{MJ1}^{\prime
}\boldsymbol{\lambda}  _{J12}^{\prime\prime\ast}\right|  $} & $\lesssim
10^{-26}\tilde{u}_{L,J}^{2}(\delta_{J}^{U})^{-1}$ & \multicolumn{1}{|c}{$23$}
& $29$ & \multicolumn{1}{l}{$29$} & \multicolumn{1}{|c}{$\mathbf{18}$} & $23$
& $23$ & \multicolumn{1}{|c}{$\mathbf{21}$} & $27$ & $27$ &
\multicolumn{1}{|c}{$\mathbf{16}$} & $\mathbf{22}$ & $\mathbf{22}$\\\hline
$\left|  \boldsymbol{\lambda}  _{212}\boldsymbol{\lambda}  _{112}%
^{\prime\prime\ast}\right|  $ & $<10^{-20}\,(\tilde{m}\sim1\,$TeV$)\;\;$ &
\multicolumn{1}{|c}{$21$} & $27$ & \multicolumn{1}{l}{$30$} &
\multicolumn{1}{|c}{$\mathbf{19}$} & $23$ & $26$ & \multicolumn{1}{|c}{$20$} &
$26$ & $28$ & \multicolumn{1}{|c}{$\mathbf{18}$} & $22$ & $25$\\
$\left|  \boldsymbol{\lambda}  _{322}\boldsymbol{\lambda}  _{112}%
^{\prime\prime\ast}\right|  $ & $<10^{-20}\,(\tilde{m}\sim1\,$TeV$)$ &
\multicolumn{1}{|c}{$23$} & $29$ & \multicolumn{1}{l}{$31$} &
\multicolumn{1}{|c}{$21$} & $25$ & $28$ & \multicolumn{1}{|c}{$21$} & $27$ &
$29$ & \multicolumn{1}{|c}{$\mathbf{19}$} & $23$ & $26$\\
$\left|  \boldsymbol{\lambda}  _{133}\boldsymbol{\lambda}  _{112}%
^{\prime\prime\ast}\right|  $ & $<10^{-21}\,(\tilde{m}\sim1\,$TeV$)$ &
\multicolumn{1}{|c}{$20$} & $26$ & \multicolumn{1}{l}{$28$} &
\multicolumn{1}{|c}{$\mathbf{18}$} & $22$ & $25$ &
\multicolumn{1}{|c}{$\mathbf{19}$} & $25$ & $27$ &
\multicolumn{1}{|c}{$\mathbf{17}$} & $21$ & $24$\\
$\left|  \boldsymbol{\lambda}  _{323}\boldsymbol{\lambda}  _{112}%
^{\prime\prime\ast}\right|  $ & $<10^{-21}\,(\tilde{m}\sim1\,$TeV$)$ &
\multicolumn{1}{|c}{$22$} & $27$ & \multicolumn{1}{l}{$30$} &
\multicolumn{1}{|c}{$\mathbf{20}$} & $24$ & $27$ &
\multicolumn{1}{|c}{$\mathbf{20}$} & $25$ & $28$ &
\multicolumn{1}{|c}{$\mathbf{18}$} & $22$ & $25$\\\hline
\multicolumn{1}{l}{$\left|  \boldsymbol{\lambda}  _{112}^{\prime\prime
}\boldsymbol{\mu}  ^{\prime I}/\mu\right|  $} & $\lesssim10^{-23}%
\tilde{u}_{R}^{2}$ & \multicolumn{1}{|l}{$22$} & \multicolumn{1}{l}{$27$} &
\multicolumn{1}{l}{$30$} & \multicolumn{1}{|l}{$\mathbf{19}$} &
\multicolumn{1}{l}{$23$} & \multicolumn{1}{l}{$26$} & \multicolumn{1}{|l}{$20$%
} & \multicolumn{1}{l}{$26$} & \multicolumn{1}{l}{$29$} &
\multicolumn{1}{|l}{$\mathbf{17}$} & \multicolumn{1}{l}{$21$} &
\multicolumn{1}{l}{$24$}\\
\multicolumn{1}{l}{$\left|  \boldsymbol{\lambda}  _{312}^{\prime\prime
}\boldsymbol{\mu}  ^{\prime I}/\mu\right|  $} & $\lesssim10^{-16}\tilde{d}%
_{R}^{2}$ & \multicolumn{1}{|c}{$18$} & $23$ & \multicolumn{1}{l}{$23$} &
\multicolumn{1}{|c}{$\mathbf{14}$} & $18$ & $18$ & \multicolumn{1}{|c}{$16$} &
$22$ & $22$ & \multicolumn{1}{|c}{$\mathbf{13}$} & $17$ & $17$\\\hline
\end{tabular}
\caption{The MFV predictions for the combinations of couplings bounded from
$\Delta B=1$ nucleon decays ($I,J=1,2,3$, $M=1,2$). The approximate bounds are
discussed in the text. For the MFV predictions, entries are to be understood
as $x\equiv\mathcal{O}(10^{-x})$. The scenarios A, B, C correspond to imposing
$SU(3)^{5}$, $SU(3)^{5}\times U(1)_{D}\times U(1)_{E}$ and $SU(3)^{5}\times
U(1)_{U}\times U(1)_{D}\times U(1)_{E}$, respectively. Finally, entries in
bold are those missing the bounds by too many orders of magnitude to be
compensated by making the squark as heavy as a few TeV.}%
\label{TableBound}%
\end{table}

\paragraph{Bounds from $\Delta B=1$ nucleon decays:}

Each bound arises from a particular mechanism and final state, hence has its
specific dependence on the intermediate sparticle masses. The strongest
bounds, taken from Ref.\cite{BarbierEtAl04,Vissani95}, are listed in Table
\ref{TableBound}. The numbers quoted for the MFV predictions are the maximum
values attainable when the indices $I,J$ ($M$) run over the three (first two)
generations, respectively. Also, for processes involving external $u_{L}$
quarks, the rotation to the mass-eigenstate basis is understood, i.e.
$u_{L}^{\prime}=Vu_{L}$ according to Eq.(\ref{Background}). For example, the
first bound in Table \ref{TableBound} stands for%
\begin{equation}
|\boldsymbol{\lambda}  _{JKI}^{\prime}\boldsymbol{\lambda}  _{11I}%
^{\prime\prime\ast}V^{\dagger K1}|
\end{equation}
For leptons, no rotation is needed since neutrino flavors are not detected.

The first three bounds in Table \ref{TableBound} correspond to tree-level
squark exchanges, and the factors%
\begin{equation}
\tilde{q}_{i}^{2}\equiv(m_{i}\,/\,100\,\text{GeV})^{2}\;,
\end{equation}
keep track of the mass suppressions. The following four make use of one $LR$
mass-insertion,%
\begin{equation}
\delta_{J}^{X}\equiv(m_{X}^{2})_{LR}^{JJ}\,/\,(m_{X}^{2})_{R}^{J}\;.
\end{equation}
At moderate $\tan\beta$, the $\mathbf{A}_{d}$ trilinear term is suppressed by
the down-type Yukawa, hence the bounds in the fourth to sixth rows are
presumably less strict than indicated. On the other hand, the bound in the
seventh row is more dangerous, both because of the potentially large stop
left-right mixing, and because $\left|
%TCIMACRO{\TeXButton{lambda}{\bm{\lambda}}}%
%BeginExpansion
\bm{\lambda}%
%EndExpansion
_{MJ1}^{\prime}%
%TCIMACRO{\TeXButton{lambda}{\bm{\lambda}}}%
%BeginExpansion
\bm{\lambda}%
%EndExpansion
_{J12}^{\prime\prime\ast}\right|  $ involves the $%
%TCIMACRO{\TeXButton{lambda}{\bm{\lambda}}}%
%BeginExpansion
\bm{\lambda}%
%EndExpansion
_{312}^{\prime\prime}$ coupling (see Table 4).

At the loop-level, all combinations of $%
%TCIMACRO{\TeXButton{lambda}{\bm{\lambda}}}%
%BeginExpansion
\bm{\lambda}%
%EndExpansion
^{\prime}$ and $%
%TCIMACRO{\TeXButton{lambda}{\bm{\lambda}}}%
%BeginExpansion
\bm{\lambda}%
%EndExpansion
^{\prime\prime}$ couplings become constrained, such that
conservatively\cite{SmirnovV96},%
\begin{equation}
\left|
%TCIMACRO{\TeXButton{lambda}{\bm{\lambda}}}%
%BeginExpansion
\bm{\lambda}%
%EndExpansion
_{IJK}^{\prime}%
%TCIMACRO{\TeXButton{lambda}{\bm{\lambda}}}%
%BeginExpansion
\bm{\lambda}%
%EndExpansion
_{I^{\prime}J^{\prime}K^{\prime}}^{\prime\prime\ast}\right|  <\mathcal{O}%
(10^{-9}-10^{-11})\;.
\end{equation}
Since $|%
%TCIMACRO{\TeXButton{lambda}{\bm{\lambda}}}%
%BeginExpansion
\bm{\lambda}%
%EndExpansion
^{\prime}|\lesssim\mathcal{O}(10^{-13})$, this bound is automatically satisfied.

There are also bounds involving $%
%TCIMACRO{\TeXButton{lambda}{\bm{\lambda}}}%
%BeginExpansion
\bm{\lambda}%
%EndExpansion
$ or $%
%TCIMACRO{\TeXButton{mu}{\bm{\mu}}}%
%BeginExpansion
\bm{\mu}%
%EndExpansion
^{\prime}$ with $%
%TCIMACRO{\TeXButton{lambda}{\bm{\lambda}}}%
%BeginExpansion
\bm{\lambda}%
%EndExpansion
^{\prime\prime}$, the strongest are from Ref.\cite{BhattacharyyaP99} and
\cite{BhattacharyyaP98}, respectively. In Table \ref{TableBound} we quote only
those which are not automatically satisfied under the MFV hypothesis.

\paragraph{Bounds from $n-\bar{n}$ oscillations:}

The neutron-antineutron oscillations set constraints on the $%
%TCIMACRO{\TeXButton{lambda}{\bm{\lambda}}}%
%BeginExpansion
\bm{\lambda}%
%EndExpansion
_{11I}^{\prime\prime}$ couplings at tree-level\cite{Zwirner83}%
\begin{equation}
\left|
%TCIMACRO{\TeXButton{lambda}{\bm{\lambda}}}%
%BeginExpansion
\bm{\lambda}%
%EndExpansion
_{11I}^{\prime\prime}\right|  \lesssim\left(  10^{-8}-10^{-7}\right)
\frac{10^{8}s}{\tau_{osc}}\left(  \frac{\tilde{m}}{100\,\text{GeV}}\right)
^{5/2}\;. \label{Bnn1}%
\end{equation}
As commented in Ref.\cite{BarbierEtAl04}, these bounds are only indicative
since the suppressions coming from $LR$ mass-insertions were ignored. This is
especially true in MFV, which tends to strongly suppress such mass insertions.
At loop-level, $n-\bar{n}$ oscillations are constraining the largest element
of $%
%TCIMACRO{\TeXButton{lambda}{\bm{\lambda}}}%
%BeginExpansion
\bm{\lambda}%
%EndExpansion
^{\prime\prime}$\cite{ChangK96}:%
\begin{equation}
\left|
%TCIMACRO{\TeXButton{lambda}{\bm{\lambda}}}%
%BeginExpansion
\bm{\lambda}%
%EndExpansion
_{312}^{\prime\prime}\right|  \lesssim\lbrack10^{-3},10^{-2}]\left(
\frac{200\text{ MeV}}{m_{s}}\right)  \;\;\;\text{ for }m_{\tilde{q}}%
\sim\lbrack100\text{ GeV, }200\text{ GeV}]\;. \label{Bnn2}%
\end{equation}
However, this bound is rather weak for squark masses above $500$ GeV,
especially compared to the one in the seventh row of Table \ref{TableBound}.%

%TCIMACRO{\FRAME{ftFU}{6.5017in}{2.1586in}{0pt}{\Qcb{The order of magnitude
%predicted by MFV for $\left|  \TeXButton{lambda}{\bm{\lambda}}_{M1I}^{\prime
%}\TeXButton{lambda}{\bm{\lambda}}_{12I}^{\prime\prime\ast}\right|  $, as a
%function of $\tan\beta$ and $m_{\nu}$ (numbers on the curves stand for
%$x\equiv\mathcal{O}(10^{-x})$). The behavior of the other bounds is similar.
%The Case I, II, IV and III correspond to the corners of the plot, starting
%from the upper left, in the clockwise direction. The three plots correspond to
%the scenarios $A,B$ and $C$ of Table \ref{TableBound}, from left to right. The
%dependence on $M_{R}$ is subleading, and we set $M_{R}=10^{-12}$%
%.\vspace{-0.5cm} }}{\Qlb{Fig1}}{Fig1.eps}%
%maintain-aspect-ratio TRUE;  display "ICON";  valid_file "F";
%width 6.5017in;  height 2.1586in;  depth 0pt;  original-width 7.5844in;
%original-height 2.4993in;  cropleft "0";  croptop "1";  cropright "1";
%cropbottom "0";  filename 'Fig1.eps';file-properties "XNPEU";}}}%
%BeginExpansion
\begin{figure}
[t]
\begin{center}
\includegraphics[
height=2.1586in,
width=6.5017in
]%
{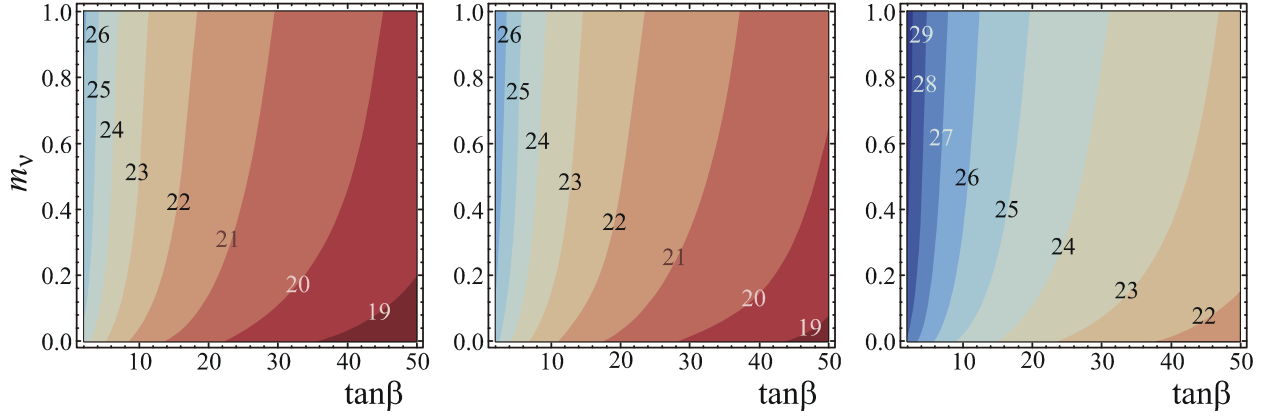}%
\caption{The order of magnitude predicted by MFV for $\left|  \bm{\lambda
}_{M1I}^{\prime}\bm{\lambda}_{12I}^{\prime\prime\ast}\right|  $, as a function
of $\tan\beta$ and $m_{\nu}$ (numbers on the curves stand for $x\equiv
\mathcal{O}(10^{-x})$). The behavior of the other bounds is similar. The Case
I, II, IV and III correspond to the corners of the plot, starting from the
upper left, in the clockwise direction. The three plots correspond to the
scenarios $A,B$ and $C$ of Table \ref{TableBound}, from left to right. The
dependence on $M_{R}$ is subleading, and we set $M_{R}=10^{-12}$%
.\vspace{-0.5cm} }%
\label{Fig1}%
\end{center}
\end{figure}
%EndExpansion

\paragraph{Conservative upper bounds:}

The order of magnitude for the combinations of couplings quoted in Table
\ref{TableBound} should be understood as conservative upper bounds for several
reasons. First, the MFV coefficients were assumed to be $\mathcal{O}(1)$. It
would however still be natural to take them of the order of the Cabibbo angle,
or suppressed by $1/4\pi$ loop-factors, leading to additional suppressions by
one or two orders of magnitude.

Secondly, throughout this work, the spurions are frozen at their background
values at a very low scale, since the light-quark masses at about $2$ GeV were
used. If we had first run these masses to the electroweak scale or higher, the
hierarchies within each coupling would have been significantly stronger since
light-fermion masses decrease rather quickly with the
energy\cite{FusaokaKoide97}. Further, the strong MFV suppressions occurring
for the couplings in Table \ref{TableBound} precisely come from light-fermion
mass factors.

The best strategy would probably have been to freeze the spurions at the scale
at which the physics leading to the MFV symmetry is thought to be acting,
presumably around the GUT scale. Then, the $RPV$ couplings could have been run
down to the electroweak scale. Such a study goes beyond the present
order-of-magnitude analysis. Anyway, it is reasonable to expect that the
running of the $RPV$ couplings is smoother than those of the light-fermion
masses, thus the bounds would again have been easier to satisfy.

Finally, we never sum over the $I,J,M$ indices, but rather scan for the
largest value. We are thus disregarding the possibility of GIM-like
cancellations. Also, we work under the assumption that only one mechanism at a
time is relevant, i.e. possible cancellations between the various processes
for a given final states are neglected.

\paragraph{Behavior in terms of $\tan\beta$ and $U(1)_{U}$, $U(1)_{D}$:}

All the couplings scale at least linearly with $\tan\beta$, so the
combinations relevant for proton decay scale at least quadratically. Overall,
the bounds become difficult to satisfy if $\tan\beta$ is too large, even with
squark masses well above $1$ TeV, and thus always require to impose either
$U(1)_{D}$ or $U(1)_{D}\times U(1)_{U}$. Note also that if $\tan\beta$ is
large and squarks are light, bounds from $\bar{n}-n$ oscillations are asking
for $U(1)_{D}$, no matter the possible suppression one could further impose on
the $\Delta L=1$ sector. In particular, the $%
%TCIMACRO{\TeXButton{lambda}{\bm{\lambda}}}%
%BeginExpansion
\bm{\lambda}%
%EndExpansion
_{1}^{\prime\prime}$ structure, with its large $312$ entry, cannot satisfy
Eq.(\ref{Bnn2}) at large $\tan\beta$ when $m_{\tilde{q}}\lesssim200$ GeV. Away
from these extreme situations, $n-\bar{n}$ oscillations are not very constraining.

If enforced, the $U(1)_{U}$ and $U(1)_{D}$ symmetries make the $\varepsilon
$-terms of $\mathbf{A}_{u}$ and $\mathbf{A}_{d}$ negligibly small, since they
are suppressed by $\det(\mathbf{Y}_{u})$ and $\det(\mathbf{Y}_{d})$.
Therefore, probing for these structures allows to test the directions in which
$B$ and $L$ are violated. For example, if the presence of the $\varepsilon
$-term in $\mathbf{A}_{d}$ is established at large $\tan\beta$, thus requiring
$U(1)_{D}$ breaking, MFV would then have difficulties with $\bar{n}-n$ oscillations.

\paragraph{Behavior in terms of $m_{\nu}$, $M_{R}$ and LFV effects:}

In Fig.1, we plot the order of magnitude predicted by MFV for $\left|
%TCIMACRO{\TeXButton{lambda}{\bm{\lambda}}}%
%BeginExpansion
\bm{\lambda}%
%EndExpansion
_{M1I}^{\prime}%
%TCIMACRO{\TeXButton{lambda}{\bm{\lambda}}}%
%BeginExpansion
\bm{\lambda}%
%EndExpansion
_{12I}^{\prime\prime\ast}\right|  $, as a function of $\tan\beta$ and $m_{\nu
}$. The behavior of the other bounds is similar. This plot permits to
interpolate between the various scenarios of Table \ref{TableBound}. The Case
I, II, IV and III correspond to the corners of the plot, starting from the
upper left, in the clockwise direction. The three plots correspond to imposing
$SU(3)^{5}$, $SU(3)^{5}\times U(1)_{D}\times U(1)_{E}$ and $SU(3)^{5}\times
U(1)_{U}\times U(1)_{D}\times U(1)_{E}$, from left to right. The dependence on
$M_{R}$ is subleading, apart from restricting $m_{\nu}$ through the
perturbative bound of Eq.(\ref{UB}). Therefore, we took $M_{R}=10^{-12}$ in
Fig.1, which allows $m_{\nu}$ to reach about $1$ eV.

The behavior of the bounds in terms of $m_{\nu}$ shown in Fig.1 can be simply
understood from the behavior of the off-diagonal elements of $\mathbf{\Upsilon
}_{\nu}$, see Eqs.(\ref{SpurionNu}) and (\ref{SpurionNu2}). Interestingly, the
size of $\Delta L=1$ couplings quickly decreases as $m_{\nu}$ increases from
zero to about $0.1$ eV. Therefore, in scenarios with large $\tan\beta$, it is
better to have $m_{\nu}$ also quite large, somewhere in the range $0.1-1$ eV.

This then has some impacts on LFV effects, and thus on rare leptonic processes
like $\mu\rightarrow e\gamma$ or $\mu\rightarrow eee$. Indeed, generically,
the bounds prefer large $m_{\nu}$, which, given Eq.(\ref{UB}), can be attained
only for small $M_{R}$. In that situation, the spurion $\mathbf{Y}_{\nu
}^{\dagger}\mathbf{Y}_{\nu}$ which tunes LFV effects (see Eq.(\ref{RPCsoft}))
is significantly suppressed.

Finally, imposing $U(1)_{L}$ would suppress all the $\Delta L=1$ couplings by
$\det(\mathbf{Y}_{\ell})$ and make it trivial to satisfy all the bounds of
Table \ref{TableBound} no matter the scenario. In that case, the only
significant constraints come from $\bar{n}-n$ oscillations, and LFV effects
can reach their maximum.

\paragraph{What to expect from $\Delta B=1$ interactions at colliders and in FCNC's:}

Except for nucleon decay, which, incidentally, may be just around the corner,
lepton-number can be considered as effectively conserved. On the other hand,
purely $\Delta B=1$ processes are not so suppressed and could offer
competitive signals in the search for supersymmetry. There exists an extensive
literature on this subject, and we do not intend to review it here (see e.g.
Ref.\cite{BarbierEtAl04}). Instead, we discuss some selected effects of
$\Delta B=1$ $RPV$ interactions, first for low-energy observables, and then at
hadron colliders.

At low energy, observables are necessarily $RPC$, hence quadratic in $%
%TCIMACRO{\TeXButton{lambda}{\bm{\lambda}}}%
%BeginExpansion
\bm{\lambda}%
%EndExpansion
^{\prime\prime}$. In particular, the tree-level squark exchanges essentially
correspond to diquark currents, with strength given by%
\begin{equation}
\frac{|%
%TCIMACRO{\TeXButton{lambda}{\bm{\lambda}}}%
%BeginExpansion
\bm{\lambda}%
%EndExpansion
_{IJK}^{\prime\prime}%
%TCIMACRO{\TeXButton{lambda}{\bm{\lambda}}}%
%BeginExpansion
\bm{\lambda}%
%EndExpansion
_{LMN}^{\prime\prime}|}{m_{\tilde{q}}^{2}}\lesssim10^{-8}\,\text{GeV}%
^{-2}\times\frac{(100\,\text{GeV})^{2}}{m_{\tilde{q}}^{2}}\;.
\end{equation}
In the favorable situation of $\tan\beta$ not too large, so that $U(1)_{D}$
does not need to be imposed, the maximum is attained within MFV for the stop,
$|%
%TCIMACRO{\TeXButton{lambda}{\bm{\lambda}}}%
%BeginExpansion
\bm{\lambda}%
%EndExpansion
_{312}^{\prime\prime}%
%TCIMACRO{\TeXButton{lambda}{\bm{\lambda}}}%
%BeginExpansion
\bm{\lambda}%
%EndExpansion
_{331}^{\prime\prime}|\sim10^{-4}-10^{-5}$ and $|%
%TCIMACRO{\TeXButton{lambda}{\bm{\lambda}}}%
%BeginExpansion
\bm{\lambda}%
%EndExpansion
_{312}^{\prime\prime}%
%TCIMACRO{\TeXButton{lambda}{\bm{\lambda}}}%
%BeginExpansion
\bm{\lambda}%
%EndExpansion
_{323}^{\prime\prime}|\sim10^{-5}-10^{-6}$. Many works have analyzed the
possible impact of these currents, for example in hadronic $B$
decays\cite{HadB}, $b\rightarrow s\gamma$\cite{bsg}, $D-\bar{D}$ and
$B-\bar{B}$ mixing\cite{DDBB} and $\Delta m_{K}$, $\varepsilon_{K}$\cite{KK}.
Not so surprisingly given that the SM contributions are tuned by $G_{F}%
\sim10^{-5}\,$GeV$^{-2}$, the bounds obtained in these works are typically $|%
%TCIMACRO{\TeXButton{lambda}{\bm{\lambda}}}%
%BeginExpansion
\bm{\lambda}%
%EndExpansion
_{IJK}^{\prime\prime}%
%TCIMACRO{\TeXButton{lambda}{\bm{\lambda}}}%
%BeginExpansion
\bm{\lambda}%
%EndExpansion
_{LMN}^{\prime\prime}|\lesssim10^{-2}-10^{-3}$ for $m_{\tilde{q}}\sim
100\,$GeV. Therefore, if MFV correctly predicts the order of magnitude of
$RPV$ effects, to have any hope to see them in low-energy $K$, $D$ or $B$
physics, the precision needed is rather challenging. Besides the experimental
difficulties, tree-level $RPV$ effects occur in hadronic channels only, whose
accuracy is ultimately limited by QCD effects.

At hadron colliders, $\Delta B=1$, $RPV$ interactions would be easier to find
because they can drastically change the phenomenology\cite{DimopoulosHall88}
(see also Refs.\cite{TEVATRON,BarbierEtAl04}).

First, the LSP can decay, mostly through hadronic channels\cite{LSP}, and
since it is no longer stable, it can be colored and/or charged. If the
neutralino is still the LSP, to identify the presence of $RPV$ would require
it to decay sufficiently quickly, within the detector. Looking back at Table
\ref{TRPV3}, one can see that the largest $%
%TCIMACRO{\TeXButton{lambda}{\bm{\lambda}}}%
%BeginExpansion
\bm{\lambda}%
%EndExpansion
^{\prime\prime}$ elements varies in the rather large range $10^{-5}\lesssim|%
%TCIMACRO{\TeXButton{lambda}{\bm{\lambda}}}%
%BeginExpansion
\bm{\lambda}%
%EndExpansion
^{\prime\prime}|_{\max}\lesssim10^{-1}$, depending on $\tan\beta$, and on the
$U(1)_{D}$ and $U(1)_{U}$ symmetries. Then, depending also on the neutralino
mass, the LSP may be effectively stable for the LHC, or may decay very
quickly. We refer to Refs.\cite{LSP} for quantitative analyses and
descriptions of the decay channels.

Secondly, single squark resonant production can occur, lowering the threshold
for the discovery of supersymmetry. With the MFV prediction $%
%TCIMACRO{\TeXButton{lambda}{\bm{\lambda}}}%
%BeginExpansion
\bm{\lambda}%
%EndExpansion
_{312}^{\prime\prime}\lesssim10^{-1}$, one would expect mostly single stop
production, which can have very distinctive signatures (see Ref.\cite{Stop}).
Also, the single gluino production through $pp\rightarrow\tilde{t}\rightarrow
t\tilde{g}$ was advocated in Ref.\cite{SopGluino} as a particularly clean
channel in which to look for supersymmetric effect at the LHC. Similarly, the
presence of the $\Delta B=1$ couplings may be felt in top-quark
production\cite{TopProd} or decay\cite{TopDecay}, which are also tuned by the
dominant $%
%TCIMACRO{\TeXButton{lambda}{\bm{\lambda}}}%
%BeginExpansion
\bm{\lambda}%
%EndExpansion
_{312}^{\prime\prime}$ coupling.

\section{Conclusion}

In this paper, we have shown that imposing the MFV hypothesis can be
sufficient to stabilize the proton. Further, MFV turns out to be more powerful
than R-parity, since it also suppresses the dangerous baryon and lepton-number
violating higher-dimensional operators. This symmetry principle, originally
introduced to solve the MSSM flavor problem, can thus successfully replace
R-parity for building a viable model. In the GUT context, or when
investigating supersymmetry-breaking mechanisms, ensuring the MFV criterium is
satisfied below the TeV scale is a simple first step towards satisfying
low-energy constraints. In this respect, depending on $\tan\beta$ and $m_{\nu
}$, one may need to allow the breaking of baryon and lepton numbers in only a
few selected directions, by restricting the number of broken flavor $U(1)$'s.

Interestingly, the MFV suppression is not always sufficient to avoid a too
rapid proton decay. This means first that depending on the values of the
parameters, proton decay and/or $n-\bar{n}$ oscillations can be very close to
their current experimental bounds. Secondly, imposing these bounds gives
indirect constraints on parameters relevant also for FCNC or LFV. Indeed,
moderate $\tan\beta$ and large $m_{\nu}$ (or alternatively the inverted
neutrino mass spectrum) are preferred, and are even compulsory if all the
$U(1)$'s are broken. On the other hand, the seesaw scale $M_{R}$ plays only a
subleading role.

MFV predicts that lepton-number can be considered as conserved in most cases,
but not baryon-number. The best signals of R-parity violation, or even of
supersymmetry, are then expected at colliders. Indeed, the impact on
low-energy observables is generically small compared to SM contributions.
Further, at tree-level, $\Delta B=1$ effects contribute dominantly to hadronic
processes, in which QCD uncertainties are quite challenging. On the other
hand, at colliders, one could look for the decays of the lightest
supersymmetric particle, not necessarily colorless and neutral. Also, MFV
predicts significant couplings for resonant stop production, as well as for
top production from down squarks. However, it should be noted that these
couplings strongly depend on the $U(1)$'s enforced, with the favorable
situation being $\lambda_{312}^{\prime\prime}\sim10^{-1}$ when all of them are
broken. Alternatively, if $U(1)_{D}$ is exact, $\Delta B=1$ couplings are all
less than about $10^{-5}-10^{-3}$, depending on $\tan\beta$, and thus may not
be readily accessible experimentally.

Though we performed our analysis in the MSSM, the viability of MFV for proton
stability is, to some extent, model-independent. Indeed, as long as the only
$(\bar{6},1)$ spurion is related to the neutrino masses, $\mathbf{\Upsilon
}_{\nu}\sim\mathcal{O}(m_{\nu}/v_{u})$, lepton-number violating couplings
remain very suppressed. In our work, the only model-dependent spurion is
$\mathbf{Y}_{\nu}^{\dagger}\mathbf{Y}_{\nu}$, remnant of the specific type I
seesaw with right-handed neutrinos. However, its impact is limited, since it
only introduces a slight softening of the hierarchies within each $RPV$
coupling, and this only when the seesaw scale is large, so that $\mathbf{Y}%
_{\nu}\sim\mathcal{O}(1)$.

Another point is that the MSSM quark and lepton flavor groups are factorized.
Indeed, no matter the precise form of the $\Delta B=1$ and $\Delta L=1$
operators inducing proton decay, MFV will separately suppress each sector, as
for the dimension-five $QQQL$ and $UUDE$ effective interactions. Of course,
this factorization no longer holds in general in GUT theories, where leptons
and quarks can be in the same multiplet, but it may nevertheless re-emerge at
the electroweak scale. For example, in $SU(5)$, the $RPV$ coupling $\bar
{5}^{I}\bar{5}^{J}10^{K}$ inducing both $\Delta B=1$ and $\Delta L=1$
couplings can be readily parametrized in terms of the $\mathbf{Y}_{\bar{5}%
}\sim\mathbf{Y}_{d}$ and $\mathbf{Y}_{10}\sim\mathbf{Y}_{u}$ Yukawa couplings
(see e.g. Ref.\cite{SU5} for the transformation rules), and is thus not
suppressed by neutrino masses. However, the smaller flavor group
$U(3)_{\bar{5}}\times U(3)_{10}$ restricts the possible directions in which
lepton and baryon numbers are violated. If we require the flavor $U(1)_{D}\sim
U(1)_{\bar{5}}$ and $U(1)_{E}\sim U(1)_{10}$ to remain exact, the coupling
$\bar{5}^{I}\bar{5}^{J}10^{K}$ as well as the bilinear $H_{\bar{5}}\bar{5}%
^{I}$ are forbidden, and R-parity violation may then arise only after the GUT
symmetry is broken. At that stage, the non-renormalizable dimension-five $RPC$
operators may also appear, but, besides being suppressed by the GUT scale,
they should still be suppressed by the MFV principle. To quantify this MFV
suppression requires to specify the dynamics of the model, at least to some
extent, and this goes beyond our bottom-up approach. Nevertheless, as said
earlier, ensuring that the factorized $U(3)^{5}$ flavor group re-emerge at
low-energy could offer an interesting strategy to keep both $RPC$ and $RPV$
flavor-breakings in check.

Finally, in the present work, cosmological implications were not investigated.
For instance, it would be interesting to analyze how the baryon asymmetry can
survive the presence of $\Delta B=1$ interactions, with the specific strengths
predicted by MFV. Also, since the MSSM LSP is no longer stable, and given that
there is always a $\Delta B=1$ coupling larger than about $10^{-5}$, it cannot
be a viable dark matter candidate. Its nature has then to be resolved at a yet
higher scale. In these contexts, ensuring that the MFV criterium is satisfied
at low energy may offer interesting constraints on possible models.

\subsection*{Acknowledgements}

We would like to thank Gilberto Colangelo for his comments and support. This
work is partially supported by the EU contract No. MRTN-CT-2006-035482
(FLAVIAnet), and by the Schweizerischer Nationalfonds.

\appendix                                                    

\section{The reduced basis for Case V}

The reduced basis in the region of parameter-space corresponding to
\begin{equation}
\text{Case V}:\tan\beta\lesssim20,\;M_{R}\lesssim2\times10^{13}\;\text{GeV}%
,\;m_{\nu}\gtrsim0.05\;\text{eV\ ,}%
\end{equation}
is rather simple because, for these values of $M_{R}$ and $m_{\nu}$, the
spurion $\mathbf{Y}_{\nu}^{\dagger}\mathbf{Y}_{\nu}$ is subleading and never
occurs. Also, $\tan\beta$ is small enough to suppress all occurrences of
$\mathbf{Y}_{\ell}^{\dagger}\mathbf{Y}_{\ell}$, which then enters only when
needed to get a non-vanishing contraction with $\varepsilon$-tensors. On the
other hand, $\tan\beta$ is large enough to feel the effects of $\mathbf{Y}%
_{d}^{\dagger}\mathbf{Y}_{d}$, and the basis we will construct for the $\Delta
B=1$ sector is in fact valid up to large $\tan\beta\approx50$. As explained in
Section 2, the expansions for $RPV$ soft-breaking terms can be immediately
obtained from those of the supersymmetric $RPV$ terms, and will not be written
down explicitly.

It should also be noted that the basis we construct can also be useful in
other situations. For example, if one insists in setting all $\Delta B=1$ MFV
coefficients to zero, but rescales $\Delta L=1$ coefficients making them of
$\mathcal{O}(M_{R})$, the MFV expansions constructed here remain valid as long
as $\mathbf{Y}_{\nu}^{\dagger}\mathbf{Y}_{\nu}\lesssim\mathbf{Y}_{\ell
}^{\dagger}\mathbf{Y}_{\ell}$, and would then offer a systematic framework for
the phenomenological study of $\Delta L=1$ effects.

Since $%
%TCIMACRO{\TeXButton{mu}{\bm{\mu}}}%
%BeginExpansion
\bm{\mu}%
%EndExpansion
^{\prime I}=\mu\mathbf{\bar{\Upsilon}}_{\nu}^{I}$, we first work out the
relevant operators in $\mathbf{\bar{\Upsilon}}_{\nu}$, and we remain with only%
\begin{equation}%
%TCIMACRO{\TeXButton{mu}{\bm{\mu}}}%
%BeginExpansion
\bm{\mu}%
%EndExpansion
^{\prime I}=\mu\,a_{1}\,\varepsilon^{LMI}(\mathbf{\Upsilon}_{\nu}^{\dagger
}\mathbf{Y}_{\ell}^{\dagger}\mathbf{Y}_{\ell})^{LM}\;.
\end{equation}
Similarly, for the $%
%TCIMACRO{\TeXButton{lambda}{\bm{\lambda}}}%
%BeginExpansion
\bm{\lambda}%
%EndExpansion
$ coupling, only two out of the several hundred operators are dominant%
\[%
%TCIMACRO{\TeXButton{lambda}{\bm{\lambda}}}%
%BeginExpansion
\bm{\lambda}%
%EndExpansion
^{IJK}=a_{2}\varepsilon^{IJL}(\mathbf{Y}_{\ell}\mathbf{\Upsilon}_{\nu
}^{\dagger})^{KL}+a_{3}\varepsilon^{KLM}\varepsilon^{IJD}\varepsilon
^{ABC}(\mathbf{Y}_{\ell}^{\dagger})^{AL}(\mathbf{\Upsilon}_{\nu}^{\dagger
})^{BD}(\mathbf{Y}_{\ell}^{\dagger})^{CM}\;.
\]

For the $%
%TCIMACRO{\TeXButton{lambda}{\bm{\lambda}}}%
%BeginExpansion
\bm{\lambda}%
%EndExpansion
^{\prime}$ and $%
%TCIMACRO{\TeXButton{lambda}{\bm{\lambda}}}%
%BeginExpansion
\bm{\lambda}%
%EndExpansion
^{\prime\prime}$ coupling, the reduced basis depends a lot on which $U(1)$ is
imposed. Let us start with the $SU(3)^{5}$ case. After expanding
$\mathbf{A}_{d}$, we find
\begin{gather}%
%TCIMACRO{\TeXButton{lambda}{\bm{\lambda}}}%
%BeginExpansion
\bm{\lambda}%
%EndExpansion
^{\prime IJK}=\varepsilon^{LMI}(\mathbf{\Upsilon}_{\nu}^{\dagger}%
\mathbf{Y}_{\ell}^{\dagger}\mathbf{Y}_{\ell})^{LM}(\left(  \mathbf{A}%
_{d}\right)  _{1}+\left(  \mathbf{A}_{d}\right)  _{2})^{KJ}\;,\\
\left(  \mathbf{A}_{d}\right)  _{1}^{KJ}=\mathbf{Y}_{d}(a_{4}\mathbf{1}%
+a_{5}\mathbf{Y}_{u}^{\dagger}\mathbf{Y}_{u}+b_{1}\mathbf{Y}_{d}^{\dagger
}\mathbf{Y}_{d}+b_{2}\mathbf{Y}_{d}^{\dagger}\mathbf{Y}_{d}\mathbf{Y}%
_{u}^{\dagger}\mathbf{Y}_{u}+b_{3}\mathbf{Y}_{u}^{\dagger}\mathbf{Y}%
_{u}\mathbf{Y}_{d}^{\dagger}\mathbf{Y}_{d}))^{KJ}\;,\nonumber\\
\left(  \mathbf{A}_{d}\right)  _{2}^{KJ}=\varepsilon^{LMN}\varepsilon
^{KBC}(\mathbf{Y}_{d}^{\dagger})^{LB}(\mathbf{Y}_{d}^{\dagger})^{MC}%
(a_{6}\mathbf{1}+a_{7}\mathbf{Y}_{u}^{\dagger}\mathbf{Y}_{u}+b_{4}%
\mathbf{Y}_{u}^{\dagger}\mathbf{Y}_{u}\mathbf{Y}_{d}^{\dagger}\mathbf{Y}%
_{d})^{NJ}\;,\nonumber
\end{gather}
and%
\begin{align}%
%TCIMACRO{\TeXButton{lambda}{\bm{\lambda}}}%
%BeginExpansion
\bm{\lambda}%
%EndExpansion
^{\prime\prime IJK}  &  =\varepsilon^{LJK}(\mathbf{Y}_{u}(a_{8}\mathbf{1}%
+a_{9}\mathbf{Y}_{u}^{\dagger}\mathbf{Y}_{u}+b_{5}\mathbf{Y}_{d}^{\dagger
}\mathbf{Y}_{d}+b_{6}\mathbf{Y}_{d}^{\dagger}\mathbf{Y}_{d}\mathbf{Y}%
_{u}^{\dagger}\mathbf{Y}_{u})\mathbf{Y}_{d}^{\dagger})^{IL}\nonumber\\
&  \;\;\;\;+a_{10}\varepsilon^{IMN}(\mathbf{Y}_{d}\mathbf{Y}_{u}^{\dagger
})^{JM}(\mathbf{Y}_{d}\mathbf{Y}_{u}^{\dagger})^{KN}\nonumber\\
&  \;\;\;\;+\varepsilon^{LMN}(\mathbf{\mathbf{Y}}_{u}(b_{7}\mathbf{1}%
+b_{8}\mathbf{Y}_{u}^{\dagger}\mathbf{Y}_{u}))^{IL}\mathbf{(\mathbf{Y}}%
_{d}\mathbf{)}^{JM}(\mathbf{Y}_{d})^{KN}\,,\nonumber\\
&  \;\;\;\;+\varepsilon^{LMN}\varepsilon^{PJK}\varepsilon^{ABI}((a_{11}%
\mathbf{1}+b_{9}\mathbf{Y}_{d}^{\dagger}\mathbf{Y}_{d})\mathbf{\mathbf{Y}}%
_{d}^{\dagger}\mathbf{)}^{LP}\mathbf{(\mathbf{Y}}_{u}^{\dagger}\mathbf{)}%
^{MA}(\mathbf{Y}_{u}^{\dagger})^{NB}\;.
\end{align}
Altogether, we therefore have $20$ operators, out of which the $9$ $b_{i}$'s
can be dropped if $\tan\beta\lesssim5$.

If we impose $SU(3)^{5}\times U(1)_{D}\times U(1)_{E}$, the term $a_{3}$ and
the whole $\left(  \mathbf{A}_{d}\right)  _{2}^{KJ}$ structure can be dropped,
while%
\begin{align}%
%TCIMACRO{\TeXButton{lambda}{\bm{\lambda}}}%
%BeginExpansion
\bm{\lambda}%
%EndExpansion
^{\prime\prime IJK}  &  =\det(\mathbf{Y}_{d})\varepsilon^{LJK}(\mathbf{Y}%
_{u}(a_{6}\mathbf{1}+a_{7}\mathbf{Y}_{u}^{\dagger}\mathbf{Y}_{u}%
)\mathbf{Y}_{d}^{\dagger})^{IL}\nonumber\\
&  \;\;\;\;+a_{8}\varepsilon^{IMN}(\mathbf{Y}_{d}\mathbf{Y}_{u}^{\dagger
})^{JM}(\mathbf{Y}_{d}\mathbf{Y}_{u}^{\dagger})^{KN}\nonumber\\
&  \;\;\;\;+\varepsilon^{LMN}(\mathbf{\mathbf{Y}}_{u}(a_{9}\mathbf{1}%
+a_{10}\mathbf{Y}_{u}^{\dagger}\mathbf{Y}_{u}))^{IL}\mathbf{(\mathbf{Y}}%
_{d}\mathbf{)}^{JM}(\mathbf{Y}_{d})^{KN}\,\nonumber\\
&  \;\;\;\;+b_{4}\det(\mathbf{Y}_{d})\varepsilon^{LMN}\varepsilon
^{PJK}\varepsilon^{ABI}(\mathbf{\mathbf{Y}}_{d}^{\dagger}\mathbf{)}%
^{LP}\mathbf{(\mathbf{Y}}_{u}^{\dagger}\mathbf{)}^{MA}(\mathbf{Y}_{u}%
^{\dagger})^{NB}\;.
\end{align}
In this case, we remain with $13$ operators, out of which the $4$ $b_{i}$'s
can be dropped if $\tan\beta\lesssim5$.

Finally, if we impose $SU(3)^{5}\times U(1)_{U}\times U(1)_{D}\times U(1)_{E}%
$, only four operators remain for $%
%TCIMACRO{\TeXButton{lambda}{\bm{\lambda}}}%
%BeginExpansion
\bm{\lambda}%
%EndExpansion
^{\prime\prime}$:%
\begin{align}%
%TCIMACRO{\TeXButton{lambda}{\bm{\lambda}}}%
%BeginExpansion
\bm{\lambda}%
%EndExpansion
^{\prime\prime IJK}  &  =\det(\mathbf{Y}_{d})\varepsilon^{LJK}(\mathbf{Y}%
_{u}(a_{6}\mathbf{1}+a_{7}\mathbf{Y}_{u}^{\dagger}\mathbf{Y}_{u}%
)\mathbf{Y}_{d}^{\dagger})^{IL}\nonumber\\
&  \;\;\;\;+\varepsilon^{LMN}(\mathbf{\mathbf{Y}}_{u}(a_{8}\mathbf{1}%
+a_{9}\mathbf{Y}_{u}^{\dagger}\mathbf{Y}_{u}))^{IL}\mathbf{(\mathbf{Y}}%
_{d}\mathbf{)}^{JM}(\mathbf{Y}_{d})^{KN}\;,
\end{align}
and we need $11$ free parameters ($8$ when $\tan\beta\lesssim5$) to describe
all supersymmetric $RPV$ couplings.

\section{The $\varepsilon$-structures for the $RPC$ trilinear terms}

For completeness, we give here the order of magnitude of the $\varepsilon
$-terms of the $RPC$ trilinear terms, Eq.(\ref{RPCsoft}). For all three, the
basic effect is to create an inverted hierarchy compared to the usual Yukawa,
though it can compete with it only at large $\tan\beta$.

For $\mathbf{A}_{u}$, whose sensitivity to $\tan\beta$ is only through
$\mathbf{Y}_{d}^{\dagger}\mathbf{Y}_{d}$ and thus very small, we find%
\[
\mathbf{A}_{u}/A_{0}=\left(
\begin{array}
[c]{ccc}%
10^{-5} & 10^{-6} & 10^{-7}\\
10^{-3} & 10^{-2} & 10^{-4}\\
10^{-2} & 10^{-2} & 1
\end{array}
\right)  +\left(
\begin{array}
[c]{ccc}%
10^{-2} & 10^{-3} & 10^{-4}\\
10^{-5} & 10^{-5} & 10^{-6}\\
10^{-9} & 10^{-8} & 10^{-7}%
\end{array}
\right)  _{\varepsilon}\;.
\]
For $\mathbf{A}_{d}$, we already gave the result while analyzing $%
%TCIMACRO{\TeXButton{lambda}{\bm{\lambda}}}%
%BeginExpansion
\bm{\lambda}%
%EndExpansion
^{\prime}$ (which involves $\mathbf{A}_{d}^{T}$), see Table 3, but we repeat
here the result for clarity:%
\begin{align*}
&  \mathbf{A}_{d}/A_{0}\overset{\tan\beta=5}{=}\left(
\begin{array}
[c]{ccc}%
10^{-4} & 10^{-7} & 10^{-6}\\
10^{-6} & 10^{-3} & 10^{-4}\\
10^{-3} & 10^{-3} & 10^{-1}%
\end{array}
\right)  +\left(
\begin{array}
[c]{ccc}%
10^{-3} & 10^{-7} & 10^{-5}\\
10^{-8} & 10^{-5} & 10^{-6}\\
10^{-8} & 10^{-8} & 10^{-6}%
\end{array}
\right)  _{\varepsilon}\;,\\
&  \mathbf{A}_{d}/A_{0}\overset{\tan\beta=50}{=}\left(
\begin{array}
[c]{ccc}%
10^{-3} & 10^{-6} & 10^{-5}\\
10^{-5} & 10^{-2} & 10^{-3}\\
10^{-2} & 10^{-1} & 1
\end{array}
\right)  +\left(
\begin{array}
[c]{ccc}%
10^{-1} & 10^{-5} & 10^{-3}\\
10^{-6} & 10^{-3} & 10^{-4}\\
10^{-6} & 10^{-5} & 10^{-4}%
\end{array}
\right)  _{\varepsilon}\;.
\end{align*}
Finally, for $\mathbf{A}_{\ell}$, the situation is similar, though we have to
distinguish the four cases of Eq.(\ref{Cases}):%
\[
\mathbf{A}_{\ell}/A_{0}\overset{\text{Case I}}{=}\left(
\begin{array}
[c]{ccc}%
10^{-5} & 10^{-11} & 10^{-11}\\
10^{-9} & 10^{-3} & 10^{-7}\\
10^{-8} & 10^{-6} & 10^{-2}%
\end{array}
\right)  +\left(
\begin{array}
[c]{ccc}%
10^{-4} & 10^{-10} & 10^{-10}\\
10^{-12} & 10^{-6} & 10^{-10}\\
10^{-13} & 10^{-12} & 10^{-7}%
\end{array}
\right)  _{\varepsilon}\;,
\]%
\[
\mathbf{A}_{\ell}/A_{0}\overset{\text{Case II}}{=}\left(
\begin{array}
[c]{ccc}%
10^{-4} & 10^{-10} & 10^{-10}\\
10^{-8} & 10^{-2} & 10^{-6}\\
10^{-7} & 10^{-5} & 10^{-1}%
\end{array}
\right)  +\left(
\begin{array}
[c]{ccc}%
10^{-2} & 10^{-8} & 10^{-8}\\
10^{-10} & 10^{-4} & 10^{-8}\\
10^{-11} & 10^{-10} & 10^{-5}%
\end{array}
\right)  _{\varepsilon}\;,
\]%
\[
\mathbf{A}_{\ell}/A_{0}\overset{\text{Case III}}{=}\left(
\begin{array}
[c]{ccc}%
10^{-5} & 10^{-7} & 10^{-7}\\
10^{-4} & 10^{-3} & 10^{-4}\\
10^{-3} & 10^{-2} & 10^{-2}%
\end{array}
\right)  +\left(
\begin{array}
[c]{ccc}%
10^{-4} & 10^{-5} & 10^{-5}\\
10^{-8} & 10^{-6} & 10^{-7}\\
10^{-9} & 10^{-8} & 10^{-7}%
\end{array}
\right)  _{\varepsilon}\;,
\]%
\[
\mathbf{A}_{\ell}/A_{0}\overset{\text{Case IV}}{=}\left(
\begin{array}
[c]{ccc}%
10^{-4} & 10^{-6} & 10^{-6}\\
10^{-3} & 10^{-2} & 10^{-3}\\
10^{-2} & 10^{-1} & 10^{-1}%
\end{array}
\right)  +\left(
\begin{array}
[c]{ccc}%
10^{-2} & 10^{-3} & 10^{-3}\\
10^{-6} & 10^{-4} & 10^{-5}\\
10^{-7} & 10^{-6} & 10^{-5}%
\end{array}
\right)  _{\varepsilon}\;.
\]

Therefore, the main impact of these $\varepsilon$-terms is to induce a larger
$LR$ mixing for the first-generation squark and slepton. This mixing is
however quite small, even at large $\tan\beta$, and it remains to be seen if
it can be singled out experimentally.


\begin{thebibliography}{9}                                                                                                %

\bibitem {FarrarF76}G.~R.~Farrar and P.~Fayet, Phys.\ Lett.\ \textbf{B76}
(1978) 575.
%%CITATION = PHLTA,B76,575;%%

\bibitem {SeeSaw}P.~Minkowski, Phys.\ Lett.\ \textbf{B67} (1977) 421;
%%CITATION = PHLTA,B67,421;%%
M.~Gell-Mann, P.~Ramond, R.~Slansky, proceedings of the Supergravity Stony
Brook Workshop, New York, 1979; T.~Yanagida, proceedings of the Workshop on
the Baryon Number of the Universe and Unified Theories, Tsukuba, Japan, 1979;
S.~L. Glashow, Quarks and Leptons, Carg\`{e}se, 1979; R.~N.~Mohapatra and
G.~Senjanovic, Phys.\ Rev.\ \textbf{D23} (1981) 165.
%%CITATION = PHRVA,D23,165;%%

\bibitem {IbanezR91}L.~E.~Ibanez and G.~G.~Ross, Nucl.\ Phys.\ \textbf{B368}
(1992) 3.
%%CITATION = NUPHA,B368,3;%%

\bibitem {BarbierEtAl04}R.~Barbier \textit{et al.}, Phys.\ Rept.\ \textbf{420}
(2005) 1 [hep-ph/0406039].
%%CITATION = PRPLC,420,1;%%

\bibitem {DambrosioGIS02}G.~D'Ambrosio, G.~F.~Giudice, G.~Isidori and
A.~Strumia, Nucl.\ Phys.\ \textbf{B645} (2002) 155 [hep-ph/0207036].
%%CITATION = NUPHA,B645,155;%%

\bibitem {PDG}W.~M.~Yao \textit{et al.} [Particle Data Group],
J.\ Phys.\ \textbf{G33} (2006) 1.
%%CITATION = JPHGB,G33,1;%%

\bibitem {ChivukulaG87}R.~S.~Chivukula and H.~Georgi,
Phys.\ Lett.\ \textbf{B188} (1987) 99.
%%CITATION = PHLTA,B188,99;%%

\bibitem {Martin97}For an introductory review of the MSSM, see for example
S.~P.~Martin, hep-ph/9709356.
%%CITATION = HEP-PH/9709356;%%

\bibitem {PecceiQ77}R.~D.~Peccei and H.~R.~Quinn, Phys.\ Rev.\ \textbf{D16}
(1977) 1791.
%%CITATION = PHRVA,D16,1791;%%

\bibitem {CiriglianoGIW05}V.~Cirigliano, B.~Grinstein, G.~Isidori and
M.~B.~Wise, Nucl.\ Phys.\ \textbf{B728} (2005) 121 [hep-ph/0507001].
%%CITATION = NUPHA,B728,121;%%

\bibitem {Weinberg79}S.~Weinberg, Phys.\ Rev.\ Lett.\ \textbf{43} (1979)
1566.
%%CITATION = PRLTA,43,1566;%%

\bibitem {BorzumatiM86}F.~Borzumati and A.~Masiero,
Phys.\ Rev.\ Lett.\ \textbf{57} (1986) 961.
%%CITATION = PRLTA,57,961;%%

\bibitem {InPrep}G.~Colangelo, E.~Nikolidakis and C.~Smith, in preparation.

\bibitem {HallR90}L.~J.~Hall and L.~Randall, Phys.\ Rev.\ Lett.\ \textbf{65}
(1990) 2939.
%%CITATION = PRLTA,65,2939;%%

\bibitem {Pheno}G.~Isidori, F.~Mescia, P.~Paradisi, C.~Smith and S.~Trine,
JHEP \textbf{0608} (2006) 064 [hep-ph/0604074];
%%CITATION = JHEPA,0608,064;%%
W.~Altmannshofer, A.~J.~Buras and D.~Guadagnoli, hep-ph/0703200.
%%CITATION = HEP-PH/0703200;%%

\bibitem {DavidsonP06}S.~Davidson and F.~Palorini, Phys.\ Lett.\ \textbf{B642}
(2006) 72 [hep-ph/0607329].
%%CITATION = PHLTA,B642,72;%%

\bibitem {MartinV93}S.~P.~Martin and M.~T.~Vaughn,
%``Two Loop Renormalization Group Equations For Soft Supersymmetry Breaking
%Couplings,''
Phys.\ Rev.\ \textbf{D50} (1994) 2282 [hep-ph/9311340].
%%CITATION = PHRVA,D50,2282;%%

\bibitem {HallS84}L.~J.~Hall and M.~Suzuki, Nucl.\ Phys.\ \textbf{B231} (1984)
419.
%%CITATION = NUPHA,B231,419;%%

\bibitem {BanksGNN95}T.~Banks, Y.~Grossman, E.~Nardi and Y.~Nir,
Phys.\ Rev.\ \textbf{D52} (1995) 5319 [hep-ph/9505248].
%%CITATION = PHRVA,D52,5319;%%

\bibitem {DavidsonE96}S.~Davidson and J.~R.~Ellis, Phys.\ Lett.\ \textbf{B390}
(1997) 210 [hep-ph/9609451].
%%CITATION = PHLTA,B390,210;%%

\bibitem {GrossmanH98}Y.~Grossman and H.~E.~Haber, Phys.\ Rev.\ \textbf{D59}
(1999) 093008 [hep-ph/9810536].
%%CITATION = PHRVA,D59,093008;%%

\bibitem {GonzalezM07}M.~C.~Gonzalez-Garcia and M.~Maltoni,
[hep-ph]0704.1800.
%%CITATION = ARXIV:0704.1800;%%

\bibitem {Cosmo}See e.g. J.~Lesgourgues and S.~Pastor,
Phys.\ Rept.\ \textbf{429} (2006) 307 [astro-ph/0603494].
%%CITATION = PRPLC,429,307;%%

\bibitem {PascoliPY03}S.~Pascoli, S.~T.~Petcov and C.~E.~Yaguna,
Phys.\ Lett.\ \textbf{B564} (2003) 241 [hep-ph/0301095].
%%CITATION = PHLTA,B564,241;%%

\bibitem {CiriglianoIP06}V.~Cirigliano, G.~Isidori and V.~Porretti,
Nucl.\ Phys.\ \textbf{B763} (2007) 228 [hep-ph/0607068].
%%CITATION = NUPHA,B763,228;%%

\bibitem {HallRS93}L.~J.~Hall, R.~Rattazzi and U.~Sarid,
Phys.\ Rev.\ \textbf{D50} (1994) 7048 [hep-ph/9306309].
%%CITATION = PHRVA,D50,7048;%%

\bibitem {Vissani95}F.~Vissani, Phys.\ Rev.\ \textbf{D52} (1995) 4245
[hep-ph/9503227].
%%CITATION = PHRVA,D52,4245;%%

\bibitem {SmirnovV96}A.~Y.~Smirnov and F.~Vissani, Phys.\ Lett.\ \textbf{B380}
(1996) 317 [hep-ph/9601387].
%%CITATION = PHLTA,B380,317;%%

\bibitem {BhattacharyyaP99}G.~Bhattacharyya and P.~B.~Pal,
Phys.\ Rev.\ \textbf{D59} (1999) 097701 [hep-ph/9809493].
%%CITATION = PHRVA,D59,097701;%%

\bibitem {BhattacharyyaP98}G.~Bhattacharyya and P.~B.~Pal,
Phys.\ Lett.\ \textbf{B439} (1998) 81 [hep-ph/9806214].
%%CITATION = PHLTA,B439,81;%%

\bibitem {Zwirner83}F.~Zwirner, Phys.\ Lett.\ \textbf{B132} (1983) 103.
%%CITATION = PHLTA,B132,103;%%

\bibitem {ChangK96}D.~Chang and W.~Y.~Keung, Phys.\ Lett.\ \textbf{B389}
(1996) 294 [hep-ph/9608313].
%%CITATION = PHLTA,B389,294;%%

\bibitem {FusaokaKoide97}H.~Fusaoka and Y.~Koide, Phys.\ Rev.\ \textbf{D57}
(1998) 3986 [hep-ph/9712201].
%%CITATION = PHRVA,D57,3986;%%

\bibitem {HadB}C.~E.~Carlson, P.~Roy and M.~Sher, Phys.\ Lett.\ \textbf{B357}
(1995) 99 [hep-ph/9506328];
%%CITATION = PHLTA,B357,99;%%
D.~Chakraverty and D.~Choudhury, Phys.\ Rev.\ \textbf{D63} (2001) 112002
[hep-ph/0012309];
%%CITATION = PHRVA,D63,112002;%%
D.~K.~Ghosh, X.~G.~He, B.~H.~J.~McKellar and J.~Q.~J.~Shi, JHEP \textbf{0207}
(2002) 067 [hep-ph/0111106];
%%CITATION = JHEPA,0207,067;%%
Y.~D.~Yang, R.~M.~Wang and G.~R.~Lu, Phys.\ Rev.\ \textbf{D72} (2005) 015009
[hep-ph/0411211];
%%CITATION = PHRVA,D72,015009;%%
R.~Wang, G.~R.~Lu, E.~K.~Wang and Y.~D.~Yang, Eur.\ Phys.\ J.\ \textbf{C47}
(2006) 815 [hep-ph/0603088].
%%CITATION = EPHJA,C47,815;%%

\bibitem {bsg}D.~Chakraverty and D.~Choudhury, Phys.\ Rev.\ \textbf{D63}
(2001) 075009 [hep-ph/0008165].
%%CITATION = PHRVA,D63,075009;%%

\bibitem {DDBB}S.~L.~Chen, X.~G.~He, A.~Hovhannisyan and H.~C.~Tsai,
[hep-ph]0706.1100.
%%CITATION = ARXIV:0706.1100;%%

\bibitem {KK}P.~Slavich, Nucl.\ Phys.\ \textbf{B595} (2001) 33
[hep-ph/0008270].
%%CITATION = NUPHA,B595,33;%%

\bibitem {DimopoulosHall88}S.~Dimopoulos and L.~J.~Hall,
Phys.\ Lett.\ \textbf{B207} (1988) 210.
%%CITATION = PHLTA,B207,210;%%

\bibitem {TEVATRON}B.~Allanach \textit{et al.} [R parity Working Group
Collaboration], hep-ph/9906224.
%%CITATION = HEP-PH/9906224;%%

\bibitem {LSP}H.~K.~Dreiner and G.~G.~Ross, Nucl.\ Phys.\ \textbf{B365} (1991)
597;
%%CITATION = NUPHA,B365,597;%%
H.~Baer, C.~H.~Chen and X.~Tata, Phys.\ Rev.\ \textbf{D55} (1997) 1466
[hep-ph/9608221];
%%CITATION = PHRVA,D55,1466;%%
B.~C.~Allanach \textit{et al.}, JHEP \textbf{0103} (2001) 048
[hep-ph/0102173].
%%CITATION = JHEPA,0103,048;%%

\bibitem {Stop}S.~Dimopoulos, R.~Esmailzadeh, L.~J.~Hall and G.~D.~Starkman,
Phys.\ Rev.\ \textbf{D41} (1990) 2099;
%%CITATION = PHRVA,D41,2099;%%
E.~L.~Berger, B.~W.~Harris and Z.~Sullivan, Phys.\ Rev.\ Lett.\ \textbf{83}
(1999) 4472 [hep-ph/9903549];
%%CITATION = PRLTA,83,4472;%%
Phys.\ Rev. \textbf{D63} (2001) 115001 [hep-ph/0012184].
%%CITATION = PHRVA,D63,115001;%%

\bibitem {SopGluino}M.~Chaichian, K.~Huitu and Z.~H.~Yu,
Phys.\ Lett.\ \textbf{B490} (2000) 87 [hep-ph/0007220].
%%CITATION = PHLTA,B490,87;%%

\bibitem {TopProd}A.~Datta, J.~M.~Yang, B.~L.~Young and X.~Zhang,
Phys.\ Rev.\ \textbf{D56} (1997) 3107 [hep-ph/9704257];
%%CITATION = PHRVA,D56,3107;%%
R.~J.~Oakes, K.~Whisnant, J.~M.~Yang, B.~L.~Young and X.~Zhang,
Phys.\ Rev.\ \textbf{D57} (1998) 534 [hep-ph/9707477];
%%CITATION = PHRVA,D57,534;%%
P.~Chiappetta, A.~Deandrea, E.~Nagy, S.~Negroni, G.~Polesello and J.~M.~Virey,
Phys.\ Rev.\ \textbf{D61} (2000) 115008 [hep-ph/9910483].
%%CITATION = PHRVA,D61,115008;%%

\bibitem {TopDecay}J.~M.~Yang, B.~L.~Young and X.~Zhang,
Phys.\ Rev.\ \textbf{D58} (1998) 055001 [hep-ph/9705341];
%%CITATION = PHRVA,D58,055001;%%
K.~J.~Abraham, K.~Whisnant, J.~M.~Yang and B.~L.~Young,
Phys.\ Rev.\ \textbf{D63} (2001) 034011 [hep-ph/0007280];
%%CITATION = PHRVA,D63,034011;%%
G.~Eilam, A.~Gemintern, T.~Han, J.~M.~Yang and X.~Zhang,
Phys.\ Lett.\ \textbf{B510} (2001) 227 [hep-ph/0102037].
%%CITATION = PHLTA,B510,227;%%

\bibitem {SU5}B.~Grinstein, V.~Cirigliano, G.~Isidori and M.~B.~Wise,
Nucl.\ Phys.\ \textbf{B763} (2007) 35 [hep-ph/0608123].
%%CITATION = NUPHA,B763,35;%%
\end{thebibliography}
\end{document}